\documentclass[11pt]{article}
\usepackage{cite}
\usepackage{amsmath,amsfonts,amssymb}
\usepackage[small,bf,hang]{caption}
\usepackage{slashed}
\usepackage{color}
\usepackage{braket}

\usepackage{tikz-cd}

\usepackage{extarrows}
\usepackage{hyperref}

\def\hybrid{
        \topmargin -20pt
        \oddsidemargin 0pt
        \headheight 0pt \headsep 0pt
        \textwidth 6.25in 
        \textheight 9.5in 
        \marginparwidth .875in
        \parskip 5pt plus 1pt \jot = 1.5ex}

\hybrid

 \csname
@addtoreset\endcsname{equation}{section}

\def\moth{\mathsurround=0pt}
\newdimen\zo \zo=0pt

\def\tick{\leaders\hrule height 0.5ex depth 0pt \hskip 0.5pt}
\def\upboxfill{$\moth \setbox\zo\hbox{\tick}%
  \hskip 3pt\hbox to 0pt{$\tick$\hss}\hrulefill \hbox to 7.5pt{$\tick$\hss}$}

\def\dtick{\leaders\hrule height .34pt depth 0.5ex \hskip 0.5pt}
\def\downboxfill{$\moth \setbox\zo\hbox{\dtick}%
  \hskip 2pt\hbox to 0pt{$\dtick$\hss}\hrulefill \hbox to 2pt{$\dtick$\hss}$}

\def\bec{\begin{center}}
\def\ec{\end{center}}

\def\be{\begin{equation}}
\def\ee{\end{equation}}
\def\bea{\begin{eqnarray}}
\def\eea{\end{eqnarray}}
\def\ba{\begin{array}}
\def\ea{\end{array}}

\def\ket#1{|#1\rangle}
\def\bra#1{\langle#1|}

\begin{document}

\begin{titlepage}
\rightline{}
\rightline{December  2021}
\rightline{HU-EP-21/56-RTG}  
\begin{center}
\vskip 1.5cm
 {\Large \bf{ Homological Quantum Mechanics }}
\vskip 1.7cm

{\large\bf {Christoph Chiaffrino, Olaf Hohm and Allison F.~Pinto}}
\vskip 1.6cm

{\it  Institute for Physics, Humboldt University Berlin,\\
 Zum Gro\ss en Windkanal 2, D-12489 Berlin, Germany}\\
\vskip .1cm

\vskip .2cm

ohohm@physik.hu-berlin.de, chiaffrc@hu-berlin.de, apinto@physik.hu-berlin.de

\end{center}

\bigskip\bigskip
\begin{center} 
\textbf{Abstract}

\end{center} 
\begin{quote}

We provide a formulation of quantum mechanics based on the cohomology of the Batalin-Vilkovisky (BV) algebra. 
Focusing on quantum-mechanical systems without gauge symmetry we introduce a homotopy retract from 
the chain complex of the  harmonic oscillator to  finite-dimensional phase space. 
This induces a homotopy transfer from the BV algebra to the algebra of functions on phase space.
Quantum expectation values for a given operator or functional are computed by the function 
whose pullback gives a functional in the same cohomology class. 
This statement is proved  in perturbation theory 
by relating  the  perturbation lemma to  Wick's theorem. 
We test  this method by computing two-point 
functions for the harmonic oscillator for position eigenstates and coherent states. 
Finally, we derive the Unruh effect,  
 illustrating that these methods are applicable to quantum field theory.

\end{quote} 
\vfill
\setcounter{footnote}{0}
\end{titlepage}

\tableofcontents


\section{Introduction}

There are two  fundamental formulations of quantum mechanics (and, by extension, of quantum field theory).
There is the canonical formulation based on states as vectors of a Hilbert space and physical observables  
as  
self-adjoint  operators acting on this space. 
This formalism determines amplitudes from which one then computes probabilities for physical observations. 
Arguably this  is still  \textit{the} definite formulation of quantum mechanics. 
Importantly, however, there is also the Feynman path integral formulation 
in which the probability amplitude between states $|q_i;t_i\rangle $ and $\langle q_f;t_f|$ is represented as 
\be
 \langle q_f;t_f|q_i;t_i\rangle = \int {\cal D}q \, \exp\Big(\frac{i}{\hbar}S[q(t)]\Big)\;. 
\ee
Here $S$ is the classical action for the dynamical variable $q(t)$, and the integral is to be taken over 
the space of all trajectories (with fixed initial value $q_i$ and final value $q_f$). 
The path integral formulation is advantageous in many respects:  
First, starting from the classical theory encoded in the action $S$, the path integral 
directly provides the objects of physical interest, the probability amplitudes, without having to deal 
with Hilbert spaces, states, operators, etc. Second, any symmetries realized as invariances of the action $S$, 
such as spacetime Lorentz invariance, are manifestly realized in the quantum theory (in the absence of anomalies). 
In fact, much of modern quantum field theory would be unthinkable without Feynman's path integral formulation. 
The trouble is that it has turned out to be very difficult, if not impossible, to give a  mathematically 
rigorous definition of the integral over the space of all kinematically allowed trajectories. 
One may discretize the physical problem at hand, in which case the path integral reduces, after Wick rotation, 
to well-defined finite-dimensional Gaussian integrals. This has been applied with great success in lattice gauge theory, 
but a general and non-perturbative definition of Feynman's path integral remains out of reach.

In this paper we introduce  an alternative algebraic formulation of quantum mechanics based 
on the (co-)homology of the Batalin-Vilkovisky (BV) algebra \cite{Batalin:1981jr,Batalin:1984jr}. 
In this we pick up on a point that to the best of our knowledge was first made by Gwilliam \cite{GwilliamThesis} and by Gwilliam and Johnson-Freyd \cite{GwilliamJF} for finite-dimensional 
toy models. 
(See also \cite{CostelloRenormalization,Costello:2016vjw,Gwilliam:2017ses,Brunetti:2013maa,Okawa:2022sjf} for related developments.) 
We  generalize their formulation  to be  applicable   to  genuine physical theories. 
Like the path integral formulation, this approach has  the advantage of 
providing  a direct way to pass from the classical theory 
to the physical quantities of the quantum theory, without having to invoke Hilbert spaces and the like. 
However, unlike Feynman's path integral, whose rigorous definition would require an infinite-dimensional generalization  of 
calculus, the formulation presented here is algebraic, employing in particular methods from 
algebraic topology.  
It should be emphasized from the outset that we do not claim to provide a full-fledged 
alternative  formulation of quantum mechanics. Rather, at the present moment, this formulation is 
restricted to the computation of  normalized quantum expectation values with respect 
to a certain class of states.

In order to begin explaining this homological approach let us consider a one-dimensional dynamical system 
with dynamical variable $q(t)$. The space of all kinematically allowed trajectories is then the infinite-dimensional 
vector space $C^{\infty}(\mathbb{R})$ of smooth functions of one real variable. 
This space being infinite-dimensional does not cause any trouble  in classical physics, where the equations of motion 
effectively make  the problem  finite-dimensional:  a solution is uniquely determined after picking 
initial or boundary conditions, say by fixing the two values $q(t_i)$ and $\dot{q}(t_i)$ at some initial time $t_i$. 
As such, in classical physics the real space of interest is just $\mathbb{R}^2$, which can be viewed as 
the phase space of this system. In contrast, in quantum mechanics it is not sufficient to go `on-shell': 
the path integral should be taken over all functions $q(t)$ in $C^{\infty}(\mathbb{R})$, 
with the ones solving the classical equations of motion being only the dominant contribution in the limit of small $\hbar$.

At this stage  concepts from topology turn out  to be helpful, notably the notion of homotopy. 
One considers two shapes or spaces to be homotopy equivalent if one can be smoothly transformed into the other, 
as for instance a closed curve which, without tearing it, can be contracted to a point. Thus, a one-dimensional curve  
may be homotopy equivalent to a zero-dimensional point. 
Similarly, we will show that passing from the infinite-dimensional 
space $C^{\infty}(\mathbb{R})$ of kinematically allowed trajectories to the finite-dimensional phase space 
$\mathbb{R}^2$ can be viewed as what is called 
a homotopy retract. While under homotopy one  of course looses some  information, the important fact is that 
the (co-)homology is homotopy invariant, this being one of the core techniques in algebraic 
topology. Since, as will be argued here,  the cohomology encodes the objects of physical interest in quantum mechanics 
this technique allows us to circumvent the hard problem of making sense of the path integral over the infinite-dimensional space 
of all trajectories and to give a more direct homological definition of quantum expectation values.

In order to discuss our main technical results in some detail, 
we begin with the algebraic structures encoding the data of physical theories. A perturbative classical field theory 
is most directly encoded in a Lie-infinity or $L_{\infty}$ algebra. This is a differential graded generalization of a Lie algebra, 
i.e., a graded vector space $V$ equipped with a  differential $\partial$ obeying $\partial^2=0$ and a potentially 
infinite number of higher brackets or maps obeying generalized Jacobi identities \cite{Zwiebach:1992ie}, see 
\cite{Lada:1994mn,Lada:1992wc,Alexandrov:1995kv,Munster:2011ij,Hohm:2017pnh} for reviews. 
The subspace $V^0$ of degree zero 
is the `space of fields' and the subspace $V^1$ of degree one is the space in which the equations of motion live or, in BV language, the space of anti-fields (in the example above both spaces are $C^{\infty}(\mathbb{R})$). 
Spaces of higher and lower degrees then encode gauge symmetries, gauge for gauge symmetries, Noether identities, etc.~\cite{Hohm:2017pnh}. (See also \cite{Erbin:2020eyc,Koyama:2020qfb,Arvanitakis:2020rrk} for the $L_{\infty}$ formulation of effective 
field theory in terms of homotopy transfer.) 
Given that $\partial^2=0$ we can consider the cohomology of $V$: the space of $\partial$-closed vectors modulo $\partial$-exact 
vectors. The cohomology encodes the on-shell data, i.e.~the (perturbative)  solutions of the classical 
equations of motion modulo gauge transformations.  

We can now explain the notion of a homotopy retract \cite{crainic2004perturbation} (see also \cite{vallette2014algebra} 
for a pedagogical introduction to the closely related notion of homotopy transfer).  
One defines a projector from the infinite-dimensional space of fields to 
the finite-dimensional phase space, for instance by projecting a field $\phi$ to its initial and final values: 
 \be\label{p}
   p : V^0
   \rightarrow \mathbb{R}^2\;, \qquad p( \phi) = (\phi(t_i), \phi(t_f)) \;. 
 \ee
There is also an inclusion map $i: \mathbb{R}^2\rightarrow V^0$ that reconstructs the solution from the initial and final values 
of $\phi$. We have a homotopy retract if there is a homotopy 
map $h: V^1\rightarrow V^0$ 
from anti-fields (or equations of motion) to fields so that 
 \be
 \begin{split}
  p\circ i &= {\rm id}\;, \\
  i\circ p &= {\rm id}-h\circ \partial \;. 
 \end{split}
 \ee
The first relation holds by definition of $p$ and $i$, and we will show that the second relation holds upon defining $h$
in terms of the Green's function. These relations tell us that while $p$ has a `right-inverse' it does not have a `left-inverse'; 
rather, $i$ is only a left-inverse `up to homotopy'. This is 
as it should be since $p$ is a genuine projector onto the much smaller space $\mathbb{R}^2$. This space  in fact equals the 
cohomology of $V$, which is homotopy invariant. 

In order to use this homotopy retract for quantum mechanics 
we then pass over to the closely related BV algebra which 
is defined on the `dual' space of smooth functions on $V$ (or rather \textit{functionals} since  $V$
is typically infinite-dimensional), which we denote by ${\cal F}(V)$.\footnote{There is also a notion of quantum or loop $L_{\infty}$-algebra that is dual to the BV algebra \cite{Zwiebach:1992ie,Markl:1997bj}, but in 
this paper we work with the BV algebra.} 
The algebra structure is given by the graded commutative and associative product of functions together with a map 
$\Delta$, called the BV Laplacian, which obeys $\Delta^2=0$. Importantly, $\Delta$ is not a derivation of the 
product but rather is a differential of `second order'. This implies that the failure of $\Delta$ to act as a derivation on 
the product defines a new structure: the anti-bracket $\{\cdot, \cdot\}$ on functions, which is a graded Lie bracket. 
In the conventional BV formalism one defines, given an action $S$,  the homological vector field  $Q=\{S,\;\cdot\;\}$, 
which acts as a derivation on functions and obeys $Q^2=0$. From this one defines 
the BV differential 
 \be\label{BVdiffIntro}
  \delta := -i\hbar \Delta +Q \;, 
 \ee
which  satisfies  $\delta^2=0$,   
provided the action satisfies the Maurer-Cartan or master equation $-i\hbar \Delta S+ \frac{1}{2}\{S,S\}=0$. 
One also introduces an odd symplectic form from which the anti-bracket is derived 
like the Poisson bracket in standard symplectic geometry. The homological vector field $Q$ is then 
the Hamiltonian vector field for the `function' $S$, and the symplectic form is $Q$-invariant. 
The BV algebra employed  in this paper deviates from the standard   construction  in the following subtle way:
The BV differential is defined by (\ref{BVdiffIntro}), but $Q$ is \textit{not} the Hamiltonian vector field for $S$  
but rather equals $\{S,\;\cdot\;\}$ only up to boundary terms. The symplectic form would then not be invariant under $Q$, but 
in our construction this is immaterial since the symplectic form makes no appearance.  
Such  generalizations of the BV formalism were introduced by Cattaneo, Mnev and Reshetikhin in \cite{Cattaneo:2012qu,Cattaneo:2015vsa}.

The homotopy retract from $V$ to $\mathbb{R}^2$ gives rise to a homotopy  retract from the BV algebra on ${\cal F}(V)$ 
to the space of functions ${\cal F}(\mathbb{R}^2)$. The BV algebra is thus transferred  to the ordinary algebra of functions 
on $\mathbb{R}^2$ concentrated in degree zero, with no non-trivial differential left, but this still encodes the complete 
cohomology of $\delta$. Our core technical  claim is now that the functions on $\mathbb{R}^2$ in the cohomology space 
compute quantum expectation values 
as follows: Given a functional $F$ of fields whose quantum expectation value we want to compute (for instance, 
for a 2-point function one considers $F=\phi(t)\phi(s)$ for fixed times $t, s$) one  determines the 
functional $F'$ that is equal  to $F$ in cohomology (so that  $F' - F = \delta G$ 
for a suitable $G$) and that is just the pullback of a function $f$ on $\mathbb{R}^2$, i.e., $F' = f \circ p$, 
where $p$ is the projector in (\ref{p}). This function $f$ computes the following normalized quantum expectation value: 
\begin{equation}\label{THESTATEMENT}
f(x,y) = \frac{\bra{y;t_f}T(F) \ket{x;t_i}}{\braket{y ; t_f|x ; t_i}} \, , 
\end{equation}
where $T$ denotes the time ordering operator, and  $\ket{x;t}$ is the state satisfying 
$\hat{\phi}(t)\ket{x;t} = x \ket{x;t}$, and similarly for $y$.
More generally, we introduce a homological procedure to compute such normalized 
expectation values with respect to states that are any linear combination of position and momentum 
eigenstates.

In perturbation theory, one can give an effective procedure to determine the function $f$ 
using the so-called perturbation lemma of homological algebra. In this case one can prove the above claim 
by showing that the computations are equivalent  to the familiar techniques based on Wick's theorem in quantum
perturbation theory, as will be illustrated with an explicit example in sec.~4. 
Our formulation  is also related to other approaches in the literature on the homotopy algebra formulation of 
perturbation theory, see 
\cite{Kajiura2003,Doubek:2017naz,Macrelli:2019afx,Jurco:2019yfd,Arvanitakis:2019ald}. While some technical details are  similar, 
and  in particular ref.~\cite{Doubek:2017naz} has been very useful for us, the general homological  formulation presented in this paper 
turns out to be quite different, as we will discuss in more detail in the summary section.  
We will also give a formal  path integral argument supporting the above claim. 
Most intriguingly,  the  homological formulation of quantum mechanics is not restricted to 
perturbation theory, yet potentially is mathematically well-defined. 

The rest of this paper is organized as follows. In sec.~2 we introduce the needed background material on BV algebras,  
and present the main statement of the homological formulation of quantum mechanics. 
We also set the stage for our subsequent applications by giving the homotopy retract for the harmonic oscillator.
In sec.~3 we compare the homological formulation with standard quantum mechanics in perturbation theory and 
using Feynman's path integral. We then illustrate and apply the homological approach in sec.~4 by computing 
expectation values for the harmonic oscillator with respect to position eigenstates and coherent states, and 
we verify that the results agree with those of conventional quantum mechanics. 
In sec.~5 we show that this approach is applicable to quantum field theory by re-deriving the Unruh effect, 
which to the best of our knowledge provides a  new derivation of this effect. 
We close in sec.~6 with a summary and outlook. In order to keep the paper self-contained we include an appendix 
summarizing the key concepts from homological algebra.

\section{General Approach}

The goal of this section is to give the homological formulation of quantum mechanics outlined above. 
In the first subsection we define  the BV algebra and review the finite-dimensional case following  Gwilliam and Johnson-Freyd
\cite{GwilliamThesis,GwilliamJF}. We then turn in the second subsection to genuine quantum-mechanical systems and 
state the main claim about the homological computation of quantum expectations values. In the final subsection
we introduce a homotopy retract for the harmonic oscillator, as preparation for the applications in later sections.

\subsection{BV for finite-dimensional Toy Model}

\subsubsection*{BV algebra}

We begin by reviewing the BV algebra and the homological formulation for a finite-dimensional toy model, 
following sec.~3 in \cite{GwilliamThesis}. 
Thinking of an `action' function $S(x)$ of a finite number of variables $x^i$, $i=1, \ldots, N$, 
the BV formalism is  defined on the larger space of functions  $F(x,x^*)$ of $x^i$ and 
new anti-commuting variables $x^*_i$. We typically consider functions of the form 
 \be
  F(x,x^*) = \sum_{k=0}^{N} f^{i_1\ldots i_k}(x) x^*_{i_1} \ldots x^*_{i_k}\;, 
 \ee
where the coefficients are smooth functions of $x$ (for instance polynomials) and antisymmetric in $i_1\ldots i_k$. 
This expansion gives a \textit{grading} to the vector  space of functions according to the number of $x^*$: functions depending only on $x$ 
are of degree zero, functions linear in $x^*$ are of degree $-1$, etc. A useful mnemonic is to assign a ghost degree of zero to 
$x$ and a ghost degree of $-1$ to $x^*$. 
In the following we will display algebraic  relations for homogenous functions of fixed degrees, denoting the degree of $F$ by $|F|$, etc., 
and we also write $(-1)^F\equiv (-1)^{|F|}$. 
More general relations then follow by linearity. 
The multiplication of functions equips this space with a graded algebra structure with $F\cdot G=(-1)^{FG} G\cdot F$.

Since the variables $x^*_i$ anti-commute one has to be careful 
with the notion of derivative: there are left- and right-derivatives which  are defined  operationally by  the  first-order variations  
 \be
  F(x,x^*+\delta x^*)- F(x, x^*) 
  =  \frac{\partial_r F}{\partial x^*_i} \delta x^*_i =  \delta x^*_i \frac{\partial_l F}{\partial x^*_i}\;, 
 \ee
so that  in general $\frac{\partial_r F}{\partial x^*_i}$ and $\frac{\partial_l F}{\partial x^*_i}$ differ by a sign: 
$\frac{\partial_r F}{\partial x^*_i} = (-1)^{F+1} \frac{\partial_l F}{\partial x^*_i}$. We follow the convention that, 
unless explicitly indicated otherwise,  all derivatives are left derivatives. 
Also note that second-order derivatives w.r.t.~$x^*$ are antisymmetric, 
$\frac{\partial^2}{\partial x^*_i \partial x^*_j}=-\frac{\partial^2}{\partial x^*_j \partial x^*_i}$. These derivatives also obey graded Leibniz rules 
w.r.t.~the multiplication of functions: 
 \be\label{tildeLeibniz}
  \begin{split}
   \frac{\partial}{\partial x^*_i}(F\cdot G) &=  \frac{\partial F}{\partial x^*_i}\cdot G +(-1)^F F\cdot  \frac{\partial G}{\partial x^*_i}\;. 
  \end{split}
 \ee

In the next step one equips this space with a  graded Lie bracket and a differential. The Lie bracket is called 
the anti-bracket and defined like the Poisson bracket: 
 \be\label{antiBracket}
  \{ F, G\} := \frac{\partial_r F}{\partial x^*_i} \frac{\partial G}{\partial x^i} 
  - \frac{\partial F}{\partial x^i} \frac{\partial_l G}{\partial x^*_i} \;. 
 \ee
This bracket has intrinsic degree of $+1$, i.e.~$|\{ F, G\}|= |F|+|G|+1$, and is graded antisymmetric with the degrees shifted by one, i.e., 
 \be
  \{F, G\} = -(-1)^{(F+1)(G+1)} \{G,F\}\;. 
 \ee
The anti-bracket obeys the graded Jacobi identity 
 \be\label{antiJacobi}
  \{\{F,G\},H\} + (-1)^{(F+1)(G+H)}   \{\{G,H\},F\} +  (-1)^{(H+1)(F+G)}  \{\{H,F\},G\} = 0 \;, 
 \ee
and the following compatibility condition with the product of functions: 
 \be\label{Compatibility}
  \{F, GH\} = \{F, G\}H + (-1)^{(F+1)G}G\{F,H\}\;. 
 \ee 
There is a  differential  of intrinsic degree $+1$, called the BV Laplacian,  defined as 
 \be\label{BVDelta}
  \Delta  = -\frac{\partial^2}{\partial x^*_i \partial x^i}\;. 
 \ee
The BV Laplacian squares to zero, 
 \be
  \Delta^2=0\;, 
 \ee
which is a consequence of ordinary derivatives commuting while derivatives w.r.t.~$x^*$ anti-commute.  
Furthermore, the Laplacian acts as a derivation on the anti-bracket: 
 \be\label{antiLeibniz}
  \Delta \{ F,G \} = \{\Delta F, G\} +(-1)^{F+1} \{F, \Delta G\} \;. 
 \ee 
In order to prove this it is convenient to first note that $\Delta$, being a second-order differential operator, does not act as a derivation 
w.r.t.~the usual multiplication of functions, but rather the anti-bracket encodes the failure of $\Delta$ to act as a derivation: 
 \be\label{derivedANti}
  (-1)^F \{F, G\} = \Delta (FG) -\Delta F G -(-1)^F F \Delta G\;. 
 \ee
This follows quickly from the definitions (\ref{antiBracket}) and (\ref{BVDelta}). Acting  on this relation with $\Delta$, using 
$\Delta^2=0$ and this relation again, then establishes the Leibniz relation (\ref{antiLeibniz}). 
This state of affairs can be summarized by saying that the anti-bracket $\{\cdot, \cdot \}$ together with the BV Laplacian $\Delta$ form a 
differential graded Lie algebra (or an $L_{\infty}$-algebra without higher brackets). More precisely, here the grading of
 the differential graded Lie algebra is given by the above grading shifted by one, explaining the presence of factors like $F+1$ 
 in many formulas.

Let us pause here to define the notion of a BV algebra. To motivate this definition we need to establish another 
relation. Acting with the Laplacian (\ref{BVDelta}) on the product of three functions yields by a straightforward computation 
using (\ref{tildeLeibniz}) 
 \be
  \begin{split}
   \Delta(FGH) = \;& \Delta( F)  G  H +(-1)^F F\Delta (G) H +(-1)^{F+G} FG\Delta(H) \\
   &+(-1)^{F}\{F, G\}H+(-1)^{F+G}F\{G, H\} +(-1)^{(F+1)G+F}G\{F, H\}\;. 
  \end{split}
 \ee
Upon using (\ref{derivedANti}) this can be rewritten in a form that employs only the (graded) multiplication of functions 
and the BV Laplacian: 
 \be\label{secondorder}
   \begin{split}
   \Delta(FGH) = \;& - \Delta( F)  G  H  -(-1)^F F\Delta (G) H  -(-1)^{F+G} FG\Delta(H) \\
   &+\Delta(FG) H +(-1)^{F} F\Delta(GH) +(-1)^{(F+1)G}G\Delta(FH)\;. 
  \end{split}
 \ee
We can now define a {BV algebra}:\\[1.5ex]
{\bf Definition:} \\
A BV algebra  is  a vector space equipped with a graded commutative and associative 
product and a degree-(+1) map $\Delta$ that satisfies  $\Delta^2=0$ and is of second order in the sense 
of  (\ref{secondorder}). \\[1.5ex] 
Note that the above definition only refers to the product structure and the differential $\Delta$, which 
is the only unconventional  ingredient since   it does  not obey the Leibniz rule with respect to the 
product but rather  the `second-order Leibniz rule' (\ref{secondorder}). The differential graded Lie algebra structure for 
the anti-bracket is then a derived notion, with the anti-bracket being defined by (\ref{derivedANti}). 
All relations of the differential graded Lie algebra  then follow from 
the axioms of a BV algebra as does the compatibility relation (\ref{Compatibility}),  
which can be derived  from (\ref{derivedANti}) and (\ref{secondorder}).

\subsubsection*{Master Equation and Cohomology}

Returning to the application of BV algebras in physical theories we note that for any differential graded Lie algebra one can write the Maurer-Cartan equation for a vector (function) of degree one 
(which  in the above grading is degree zero). An example of such a function is the action $S$, and the Maurer-Cartan equation 
reads
\be\label{MasterEQ}
\frac{1}{2}\{ S,S\} -i\hbar \Delta S = 0\;, 
\ee
where the differential $\Delta$ was rescaled by $i\hbar$ so that this takes the form of the BV master equation. 
Given a solution $S$ of the Maurer-Cartan equation one can define a new differential $\delta$, 
 \be\label{deltaDEF}
   \delta := \{S,\,\cdot\, \}  - i\hbar \Delta \;, 
 \ee
which squares to zero, $\delta^2=0$,  if and only if the master equation (\ref{MasterEQ})  holds.  
This follows by a quick computation using $\Delta^2=0$,  the Jacobi identity (\ref{antiJacobi}) and the Leibniz rule (\ref{antiLeibniz}).\footnote{Using (\ref{Compatibility}) one may  
verify that $Q\equiv \{S,\cdot\}$ acts as a derivation with respect to the product, so that with (\ref{derivedANti}) the anti-bracket can also
be defined in terms of the full BV differential as 
 \be\label{deltaFailure}
   -i\hbar (-1)^F \{F, G\} = \delta(FG) - \delta F G - (-1)^F F\delta G \;. 
 \ee }
In the present context, an $S$ that does not depend on $x^*$ solves the master equation trivially: both terms in 
(\ref{MasterEQ}) vanish separately. Note, finally, that the master equation (\ref{MasterEQ}) is also equivalent to 
 \be\label{MasterMaster}
  \Delta \big(e^{\frac{i}{\hbar}S}\big)=0\;, 
 \ee
where the exponential of the degree-zero object $S$ is defined in terms of its Taylor  series. 
This relation can be verified  with (\ref{derivedANti}). 

Since $\delta^2=0$ there is a notion of homology spaces $\frac{{\rm ker}\,\delta}{{\rm im}\,\delta}$: spaces of $\delta$-closed 
vectors modulo $\delta$-exact vectors. Our goal is now to compute this homology for a simple toy model with one real variable $x$ 
and action 
 \be
  S(x) = \frac{1}{2} ax^2\;. 
 \ee 
We will see that the homology computes the `path integral' including $e^{\frac{i}{\hbar}S}$, which here reduce to regular integrals.

We begin by writing out the differential $\delta$ defined in (\ref{deltaDEF}), using that here only two variables $x$ and $x^*$ with $(x^*)^2=0$ enter: 
 \be
 \begin{split}
  \delta &= - \frac{\partial S}{\partial x} \frac{\partial}{\partial x^*} + i\hbar \frac{\partial^2}{\partial x\partial x^*}  \\
  &= -a x \frac{\partial}{\partial x^*} + i\hbar \frac{\partial^2}{\partial x\partial x^*}\;. 
 \end{split} 
 \ee
First, we have to determine ${\rm ker}\,\delta$, the space of functions that are $\delta$-closed. A general function can be written as 
the superfield 
 \be\label{Fexpansion}
  F(x,x^*) = f(x)+ x^* g(x)\;, 
 \ee
where as above we assume that $f$ and $g$ are polynomials. This yields 
 \be\label{deltaF}
  \delta F(x,x^*) = -a x g(x) + i \hbar g'(x) \;. 
 \ee
Setting this to zero gives  a differential equation that does not have a non-trivial solution in polynomials (otherwise one has  the 
solution $e^{ia \frac{x^2}{2\hbar}}$). Therefore, the kernel of $\delta$ is given by functions (\ref{Fexpansion}) with $g=0$: 
 \be
  {\rm ker}\,\delta = \{ F(x,x^*) \equiv f(x)\}\;. 
 \ee
Next we have to mod out by ${\rm im}\,\delta$, the space of $\delta$-exact functions. In order to see the significance of this step
let us consider functions of the form (\ref{Fexpansion}) with $f=0$ and $g=x^n$, which yields with (\ref{deltaF}) 
 \be
  \delta(x^* x^n) =  - a x^{n+1} + i\hbar n x^{n-1} \;. 
 \ee
When passing over to homology we identify any two functions that differ by a $\delta$-exact function. Hence the above relation 
implies that its right-hand side is zero in homology or, equivalently, that we have the equivalence 
 \be
  x^{n+1} \sim \frac{i\hbar}{a} n x^{n-1} \;. 
 \ee
For instance, $x^2$ is equivalent (can be reduced to) a constant, $x^2\sim \frac{i\hbar}{a}$. Similarly, $x^3$ can be reduced 
to $x$, since  $x^3\sim \frac{i\hbar}{a} 2x$, but noting that $x$ is $\delta$-exact, $x=-\frac{1}{a} \delta(x^*)$, this is actually equivalent to zero. 
Using this iteratively we have 
 \be\label{VEVxn}
  x^n \ \sim \ \left\{
  \begin{array}{l l} 0 
    & \quad \text{for $n$ odd}\\
    \left(\frac{i\hbar}{a}\right)^{\frac{n}{2}} (n-1)(n-3)\cdot \cdots \cdot 1 & \quad \text{for $n$ even }\\
  \end{array} \right. 
 \ee
The reader may recognize the right-hand side as the `quantum expectation value' $\langle x^n \rangle$ of a zero-dimensional QFT with a single 
variable (see, e.g., p.~14 in Zee's text book \cite{Zee:2003mt}). More precisely, let us define the expectation value for a polynomial $f$ in terms 
of the convergent  Gaussian integral as 
 \be
  \langle f \rangle  := \frac{1}{{\cal N}}\int_{-\infty}^{\infty} dx \,f(x) \, e^{-\frac{a x^2}{2\hbar}}\Big|_{a\rightarrow -ia }\;, 
 \ee
with normalization ${\cal N}:=\int_{-\infty}^{\infty} dx  e^{-\frac{a x^2}{2\hbar}}$, where the substitution $a\rightarrow -ia$ 
is done after performing the integral. 
Then the right-hand side of (\ref{VEVxn}) 
equals $\langle x^n\rangle$.  
More generally,  any polynomial $f$ is equivalent in homology to a complex number, and identifying its homology class $[f]$ with this 
 number we have 
  \be
   [f] = \langle f \rangle \;. 
  \ee
Thus, the homology of the differential $\delta$ computes expectation values.

\subsection{BV for Quantum Mechanics}\label{SectionStatement}

We now turn to generic dynamical systems and state our main claim about the homological formulation of the corresponding 
quantum theories. For definiteness let us consider a one-dimensional mechanical model with action
\begin{equation}
S[\phi] = \int_{t_i}^{t_f} \text d t \, L(\phi(t),\dot \phi(t),t) \, .
\end{equation}
The fields here are smooth functions $\phi$ on the interval $[t_i,t_f]$, i.e.~$\phi \in C^\infty([t_i,t_f])$. Formally, such theories include genuine (quantum) field theories, where the dependence of $\phi$ on spatial coordinates is suppressed, 
but for definiteness let us think of ordinary  quantum mechanics. 
 
We start from the BV formalism as in the previous section. This means that the space of dynamical variables is enlarged to 
include, in addition to $\phi$, anti-fields $\phi^*$, which we also assume to be smooth, $\phi^* \in C^\infty([t_i,t_f])$. Fields and anti-fields are combined into a total space
\begin{equation}\label{complexV}
V = V^0 \oplus V^{1}\;, \qquad  V^0\equiv C^{\infty}([t_i,t_f])\;, \quad V^{1}\equiv  \Pi C^{\infty}([t_i,t_f]) \, , 
\end{equation}
which we also sometimes denote by $V^{\bullet}$, to indicate that we refer to the total space of a chain complex. 
Moreover, here we used, for any vector space $X$, the common notation $\Pi X$ to indicate that the elements of $\Pi X$ have reversed parity  compared to $X$. This means, for instance,  that taking the elements $\phi\in C^{\infty}([t_i,t_f])$ to be of even degree, the elements $\phi^*\in  \Pi C^{\infty}([t_i,t_f])$ 
have odd degree, although in the end they are both just smooth functions. 
The definition of the BV algebra given in the previous subsection goes through in the present context, just with functions replaced by functionals 
and derivatives replaced by functional derivatives. Concretely, we consider functionals that are superpositions of the monomials
\begin{equation}\label{FunctionalMechanics}
F[\phi,\phi^*] = \int \text d t_1 \cdots \text d t_k \text d s_1 \cdots \text d s_l f(t_1,...,t_k,s_1,...,s_l) \phi(t_1)\cdots \phi(t_k) \phi^*(s_1) \cdots \phi^*(s_l) \, , 
\end{equation}
where the coefficient functions $f(t_1,...,t_k,s_1,...,s_l)$ are completely symmetric in the $t_i$ and completely  antisymmetric in the $s_i$.  
The degree of such a functional is minus the number of anti-fields appearing in there, i.e., $|F| = -l$. (The relative minus sign 
compared to the degree of $+1$ for $\phi^*$ encoded in (\ref{complexV}) is not a typo but rather  due to
functionals being dual to (anti-)fields, hence having the opposite degree.)
We denote this space of functionals by $\mathcal{F}(V)$, which inherits a grading by the number of $\phi^*$, so that we can write 
${\cal F}(V)=\cdots\oplus {\cal F}(V)^{-2}\oplus {\cal F}(V)^{-1} \oplus {\cal F}(V)^{0}$. The BV algebra structure is defined 
on this space as in the previous subsection. For instance, the differential $\delta = Q  - i\hbar \Delta$, which still 
satisfies  $\delta^2 = 0$, consists of two parts. There is the classical piece (surviving the limit $\hbar \rightarrow 0$) given by 
the homological vector field
\begin{equation}\label{homologicalQ}
Q = -\int_{t_i}^{t_f}\text d t \, EL(\phi(t))\frac{\delta}{\delta \phi^*(t)} \, , 
\end{equation} 
where $EL(\phi(t)) = 0$ are the Euler Lagrange equations. It also satisfies $Q^2=0$. The second piece is the BV Laplacian 
 \be
  \Delta=-\int_{t_i}^{t_f}\text d t \,\frac{\delta}{\delta \phi(t)\delta \phi^*(t)}\;. 
 \ee
Note that $Q$ and $\delta$  both decrease the number of $\phi^*$ by one and hence have an intrinsic degree of $+1$. 
(Sometimes we denote by $Q_p$ the restriction of $Q$ to ${\cal F}(V)^{p}$.)

We have to comment on the following deviation from the standard BV formalism: 
in general  $\frac{\delta S}{\delta \phi(t)}$ is not equal to $EL(\phi(t))$, and so in general   $Q \ne \{S,-\}$. 
The reason is that these two are only equal up to boundary terms. (Later it will be crucial to allow for variations along solutions, 
and these variations do not vanish on the boundary.) 
Indeed, we take the space of fields  to be the space of smooth functions defined on the interval $[t_i,t_f]$ rather than the entire  real line. This makes the action well-defined without having to assume any behavior of the fields at infinity.  
However, in contrast to standard treatments of the BV formalism, the symplectic form $\omega$ inducing the bracket $\{-,-\}$ is no longer invariant under the vector field $Q$ due to boundary terms. An extension of BV formalism, called BV-BFV formalism,  accommodates such features and 
was developed by Cattaneo, Mnev and Reshetikhin in \cite{Cattaneo:2012qu,Cattaneo:2015vsa}, 
but in this work we never use the symplectic form, so these issues are of no concern to us.

Given that $\delta^2 = 0$ we can define the cohomology of $\mathcal{F}(V)$ with respect to $\delta$, and we will show that this cohomology computes quantum expectation values. Before turning to this  it is instructive to inspect the cohomology $H^p(Q)= \frac{\ker Q_p}{\text{im} \, Q_{p-1}}$ of $\mathcal{F}(V)$ with respect to $Q$, 
recalling $Q^2=0$. Specifically, we will first establish the following \\[1.5ex]
\textit{Claim: The cohomology of $Q$ is isomorphic to the space of ``on-shell functionals".}\\[1.5ex]
Here on-shell functionals refer to functionals on solutions 
to the equation of motion $EL(\phi(t)) = 0$. 
More precisely, this holds under the assumption that there are no non-trivial gauge symmetries. 
In the following we will establish this statement. 
To this end let us introduce  some notation: We denote the subspace of solutions by ${\cal E} \subseteq V^0$,  
 \be
  \mathcal{E} = \Big\{ \phi\in V^0\Big|EL(\phi(t))=0\Big\}\;. 
 \ee
Since $\mathcal{E}$ is  a subspace of  $V^0$, any functional $F$ on $V^0$ can be restricted to a functional on $\mathcal{E}$. This defines a projection
\begin{equation}
\begin{split}
r: \mathcal{F}(V^0)&\longrightarrow \mathcal{F}(\mathcal{E}) \, , \qquad 
	F \longmapsto F|_{\mathcal E} \, .
\end{split}
\end{equation}
Let us assume that any functional on $\mathcal{E}$ is obtained in this way. In other words, any functional on ${\cal E}$ extends to a functional on $V^0$. Then the restriction map $r$ is surjective. In general, however, $r$ is not a bijection since it may have a non-trivial kernel, 
but upon modding out the kernel we have an isomorphism:  $\mathcal{F}(V^0)/{\rm ker}(r) \cong \mathcal{F}(\mathcal E)$.

We now compute the cohomologies in order to establish the above claim. 
We begin at degree zero and note that functionals of degree zero are annihilated by $Q$, hence  $\ker Q_0=\mathcal F(V^0)$. 
We next show that in degree zero the image of $Q$ consists of functionals of $\phi$ proportional to the equation of motion. 
To see this consider a functional of degree $-1$, i.e., 
$G_{-1} = \int_{t_i}^{t_f} \text d s \, \phi^*(s) g_{-1}[\phi,s]$, where  
$g_{-1}[\phi,s]=\int dt_1\cdots dt_k f_{-1}(t_1, \ldots, t_k,s)\phi(t_1)\cdots \phi(t_k)$. We then have 
\begin{equation}\label{QG}
Q(G_{-1}) = -\int_{t_i}^{t_f} \text d t \, EL(\phi(t)) g_{-1}[\phi,t] \, , 
\end{equation}
and so this functional vanishes on ${\cal E}$. Therefore, $Q(G_{-1})$ is  in the kernel of $r$, i.e.~$\text{im} \, Q_{-1} \subseteq \ker r$. 
Assuming that any functional in the kernel of $r$ is of the form (\ref{QG}) (c.f.~appendix B in \cite{Gomis:1994he} for a discussion on this assumption), for a suitable function $g_{-1}$, one has in fact 
$\text{im} \, Q_{-1} = \ker r$. We then find for the cohomology 
\begin{equation}
   H^0(Q) = \frac{\ker Q_0}{\text{im} \, Q_{-1}} = \frac{\mathcal F(V^0)}{\ker r} \cong \mathcal{F}(\mathcal E) \, .
\end{equation}
So the cohomology in degree zero is in fact equal to the space of functionals on solutions.
This establishes the claim for functionals of degree zero. 

Let us now turn to the cohomology in degree minus one. The kernel of $Q_{-1}$ consists of functionals $G_{-1}$ of the form given above (\ref{QG}) 
for which  $Q(G_{-1})=0$. There is a straightforward physical interpretation of this condition: 
inspecting the explicit form (\ref{QG}) one infers that 
$Q(G_{-1}) = 0$ iff $\delta \phi(t) \equiv g_{-1}[\phi,t]$ is a (gauge) invariance of the action. 
We next  investigate which functionals at degree $-1$ are in the image of $Q_{-2}$. 
Consider a generic functional at degree $-2$, 
 \be
  G_{-2} = \int ds_1ds_2\phi^*(s_1) \phi^*(s_2) g_{-2}[\phi,s_1,s_2]\;. 
 \ee
 Computing $Q(G_{-2})$ one obtains functionals of the form $G_{-1}$ given above (\ref{QG}) where 
\begin{equation}
g_{-1}[\phi,t] = \int_{t_i}^{t_f} \text d s \, EL( \phi(s)) g_{-2}[\phi,s,t] \, , \qquad g_{-2}[\phi,s,t] = -g_{-2}[\phi,t,s] \, . 
\end{equation}
Under the interpretation of gauge symmetries these are precisely the 
 trivial gauge symmetries, which  
always exist. Their  action vanishes once we restrict to solutions. 
We therefore learned that 
the trivial symmetries are in the image of $Q_{-2}$ and hence  that $\frac{\ker Q_{-1}}{\text{im} \, Q_{-2}}$ parametrizes non-trivial gauge symmetries. When there are no non-trivial gauge symmetries, the cohomology in degree $-1$ therefore vanishes. A similar argument applies  in the case of the cohomologies in arbitrary negative degree. This completes our discussion of the claim that the cohomology encodes ``on-shell functionals", i.e., functionals 
 on solutions of the equations of motion. 

\medskip

For the following applications it will be important to have  an explicit model for the space of solutions. Suppose that $\phi_0$ is a solution to the equation of motion. Any such solution is uniquely determined by its boundary values $x_i = \phi(t_i)$ and $x_f = \phi(t_f)$. We therefore have 
${\cal E} \cong \mathbb{R}^2$ and so $\mathcal{F}(\mathcal E) \cong \mathcal{F}(\mathbb{R}^2)$. Any functional on the space of solutions can be viewed as an ordinary function $f(x_i,x_f)$ in two variables.

The space $\mathbb{R}^2$ of boundary values naturally embeds into $V^{\bullet}$. Given $(x_i,x_f) \in \mathbb{R}^2$, let $\phi_{x_i,x_f}$ be the unique solution with these boundary conditions, i.e., we demand that
\begin{equation}\label{BoundaryCondition}
(\phi_{x_i,x_f}(t_i),\phi_{x_i,x_f}(t_f)) = (x_i,x_f) \, .
\end{equation} This defines the embedding or inclusion map 
\begin{equation}
\begin{split}
i: \mathbb{R}^2 \longrightarrow V^{\bullet} \, , \qquad 
(x_i,x_f) \longmapsto (\phi,\phi^*) = (\phi_{x_i,x_f},0) \, .
\end{split}
\end{equation}
There is also an associated projection
\begin{equation}
\begin{split}
p: V^{\bullet} \longrightarrow \mathbb{R}^2 \, , \qquad 
(\phi,\phi^*) \longmapsto (\phi(t_i),\phi(t_f)) \,,
\end{split}
\end{equation}
satisfying $p \circ i = \text{id}_{\mathbb{R}^2}$ according to \eqref{BoundaryCondition}.

Given the above maps we can define their pullbacks that act (in the opposite direction) on the dual spaces of functionals and functions. 
The pullback of the inclusion is the map
 \be
  i^{*}: {\cal F}(V^{\bullet}) \longrightarrow {\cal F}(\mathbb{R}^2)\;, \qquad i^{*}(F) := F\circ i\;. 
 \ee
Thus, this map associates to a functional ${F}$ the function on $\mathbb{R}^2$ defined by $i^{*}(F)(x_i, x_f)=F[\phi_{x_i,x_f},0]$. 
Similarly, the pullback of the projection is the map 
 \be
  p^{*}: {\cal F}(\mathbb{R}^2) \longrightarrow {\cal F}(V^{\bullet})\;, \qquad p^{*}(f) = f\circ p\;. 
 \ee
Therefore, any function $f$ defines a functional  via $F[\phi,\phi^*] = f(p([\phi,\phi^*])) = f(\phi(t_i),\phi(t_f))$.
We say that a functional $F$ restricts to $\mathbb{R}^2$, or  that $F$ is an ``on-shell functional",  
 if it is the pullback of a function on $\mathbb{R}^2$. 
In order to identify $\frac{\ker Q}{\text{im} \, Q}$ with $\mathbb{R}^2$, we can look for representatives $F$ of equivalence classes $[F] \in \frac{\ker Q}{\text{im} \, Q}$ that are of that form. Explicitly, given a functional $F[\phi]$ at degree zero, 
we can look for a $G[\phi,\phi^*]$ of degree $-1$, such that $F' = F + Q(G)$ 
is the pullback of a function $f$ on $\mathbb{R}^2$ and can hence be written as $F' = f \circ p$.

As an aside we  note that the pullback maps $i^{*}$ and $p^{*}$  define what are called chain maps in the homological language. 
When we think of $\mathcal{F}(\mathbb{R}^2)$ as a complex in degree zero with trivial cohomological vector field $\tilde{Q} = 0$, the 
maps $i^{*}$ and $p^{*}$ obey 
 \be
  i^{*}\circ Q= \tilde{Q}\circ i^{*}  =0 \;, \qquad Q\circ p^{*} = p^{*}\circ \tilde{Q}=0\;, 
 \ee
which means that 
\begin{equation}
 (QF) \circ i = 0 , \qquad Q(f\circ p) = 0 
\, .
\end{equation}
The first relation follows 
since the image of $i$ are the solutions on which $Q$ vanishes, while 
the second relation follows since $f\circ p$ is a functional of degree $0$.

We now turn to our main statement relating to quantum mechanics. 
Thinking of $-i\hbar \Delta$ as a small perturbation, we expect that the cohomology of $\delta = Q - i \hbar \Delta$ is the same as that of $Q$.   
We can then search for functionals in the $\delta$ cohomology class of  $F$ that restrict to $\mathbb{R}^2$, i.e.~we look for a functional $G$, such that
$F' = F + \delta(G)$ and $F' = p^*(f)= f \circ p$ for some function $f$ on $\mathbb{R}^2$. 
This function $f$ is unique since $p^*$ is invertible in cohomology. 
We claim that this function computes a certain normalized expectation value of $F$. More precisely,
\begin{equation}\label{THESTATEMENT}
f(x,y) = \frac{\bra{y;t_f}T(F) \ket{x;t_i}}{\braket{y ; t_f|x ; t_i}} \, .
\end{equation}
On the right-hand side, we use the language of canonical quantization, where $T$ denotes the time ordering operation. Furthermore, the state $\ket{x;t}$ is the state that satisfies the eigenvalue equation $\hat{\phi}(t)\ket{x;t} = x \ket{x;t}$, and similarly for $y$, where  $\hat \phi(t)$ is the canonically quantized position operator at time $t$ associated to the classical field $\phi(t)$.

Let us point out that while the above statement was made in the context of a particular projector mapping the infinite-dimensional vector space 
$V$ to $\mathbb{R}^2$, this relation between cohomology and quantum expectation values holds  more generally. 
For instance, we may consider the more general projectors
\begin{equation}
\begin{split}
p: V &\longrightarrow \mathbb{R}^2 \, , \\
\phi &\longmapsto (a_i \phi(t_i) + b_i \dot \phi(t_i), a_f \phi(t_f) + b_f \dot \phi(t_f)) \, .
\end{split}
\end{equation}
In this case, the boundary states change. The incoming state $\ket{x,t_i}$ is now an eigenstate of the operator $a_i\hat \phi(t_i)+ b_i \hat{\pi}(t_i)$ with eigenvalue $x$, where  $\hat{\pi}(t_i)$ is the momentum operator at $t = t_i$. Similarly, the outgoing state $\bra{y,t_f}$ is an eigenstate of $a_f\hat \phi(t_f)+ b_f \hat{\pi}(t_f)$ with eigenvalue $y$.

\subsection{Homotopy Retract for Harmonic Oscillator}

\label{ClassicalHO}

In the previous subsection we have argued that for a generic dynamical system the 
infinite-dimensional function space of dynamical variables  is equivalent in 
cohomology to the finite-dimensional phase space (or the space of initial conditions). 
Here we discuss further the topological interpretation of this fact by 
displaying, for the case of the harmonic oscillator, 
a so-called homotopy retract that maps between the corresponding 
chain complexes. This will be instrumental for the applications to be discussed below. 

The equation of motion for the harmonic oscillator is
\begin{equation}\label{HarmOsc}
EL( \phi(t)) = - (\ddot{\phi} + \omega^2 \phi) = 0 \, , 
\end{equation}
and defines the BV differential $Q$ according to (\ref{homologicalQ}). 
Since the cohomology of $Q$  describes the space of functionals on the solution space of $\ddot{\phi} + \omega^2 \phi = 0$, 
which is isomorphic to the space of initial conditions $(\phi(t_i),\dot \phi(t_i)) \in \mathbb{R}^2$, 
we expect that the cohomology of $Q$ should be isomorphic to  functions on $\mathbb{R}^2$.
Instead of working on the spaces of functionals, on which $Q$ acts as a vector field, we can 
work directly on the field space encoded in the  complex
\begin{equation}\label{HOComplex}
\begin{tikzcd}
0 \arrow[r] & V^0 \arrow[r, "\partial"] & V^{1} \arrow[r] & 0\;, 
\end{tikzcd}
\end{equation}
where we recall that $V^0$ and $V^1$ are two copies of $C^{\infty}([t_i,t_f])$. 
The non-trivial differential is $\partial(\phi) = \ddot \phi + \omega^2 \phi$.\footnote{The space of \textit{linear} functionals 
is the dual space to $V^{\bullet}$, 
and for a linear $F$ we have $(QF)[\phi,\phi^*] =F[0,\partial(\phi)]$, so that $\partial$ is the map dual to $Q$ 
(noting that $\partial(\phi^*)=0$). } 
The cohomology of this complex in degree zero is $\ker \partial$, so it consists of the space of solutions. Since any solution is determined by its initial condition $(\phi(t_i),\dot \phi(t_i))$, we can identify the cohomology with the phase space $\mathbb{R}^2$. In degree one, the cohomology is zero, because any element is $\partial$-exact: for any 
function $f(t)$ there is a $\phi(t)$ so that 
\begin{equation}\label{DrivenHarmOsc}
\partial (\phi) = \ddot{\phi} + \omega^2 \phi = f \, .
\end{equation}
From the theory of ordinary differential equations we know that a solution, which we denote by $\phi_f$, exists under very mild assumptions on $f$. Smoothness of $f$ ensures that 
$\phi_f$ is also smooth. We can give an explicit solution to \eqref{DrivenHarmOsc} 
by means of a Green's function: 
\begin{equation}\label{Propagator}
\phi_f(t) = \int_{t_i}^{t_f} \text d s\, K(t,s)\, f(s)\;.  
\end{equation}
Explicitly,  the corresponding kernel function reads 
\begin{equation}
K(t,s) := \theta(t-s)\frac{\sin \omega(t-s)}{\omega}\;, 
\end{equation}
with $\theta$ the step function, and it satisfies 
 \be\label{Kernel}
  (\partial^2_t + \omega^2)K(t,s) = \delta(t-s)\;. 
 \ee
Of course, $\phi_f(t)$ is unique only up to homogeneous solutions, i.e.~solutions to \eqref{HarmOsc}.  
We picked a solution such that $\phi_f(t_i) = \dot{\phi}_f(t_i) = 0$.

Our above analysis shows that the cohomology space  $\frac{\ker \partial}{\text{Im} \, \partial}$ is isomorphic to 
the phase space $\mathbb{R}^2$, which we can think of as a chain complex in degree zero with trivial differential. 
We have the following maps called quasi-isomorphisms: 
\begin{equation}
\begin{tikzcd}
0 \arrow[r] & V^0 \arrow[r,"\partial"] \arrow[d,"p"]& V^1 \arrow[r] \arrow[d,"0"] & 0  \\
0 \arrow[r] & \mathbb{R}^2 \arrow[r,"0"] & 0 \arrow[r] & 0
\end{tikzcd}\, 
\end{equation}
where
\begin{equation}\label{Rpro}
p(\phi) = (\phi(t_i),\dot \phi(t_i)) \, .
\end{equation}
There is also a quasi-isomorphism $i: \mathbb{R}^2 \rightarrow V^0$ in the other direction, defined by 
\begin{equation}\label{Rinc}
i(q,p) = q \cos \omega (t-t_i) + \frac{p}{\omega} \sin \omega (t-t_i) \, . 
\end{equation}
This obeys  $p \circ i = 1$.

In the following we will construct a so-called homotopy retract of the original chain complex $V^{\bullet}$ to 
its cohomology $\mathbb{R}^2$.  
To this end we interpret the Green's function (\ref{Propagator}) as a   map $h: V^{1} \rightarrow V^0$, 
i.e., for any function $f\in V^1$ we define 
\begin{equation}
h(f)(t) = \int_{t_i}^{t_f}\text d s\, K(t,s)\, f(s)\;.
\end{equation}
We can extend $h$ to the full complex $V^\bullet$ by defining $h(\phi) = 0$ for any field $\phi \in V^0$. The map  $h$ is 
called a homotopy retract from the original complex $V^\bullet$ to its cohomology $\mathbb{R}^2$. In general, we say that a complex $(Y,\partial_Y)$ is a homotopy retract of $(X,\partial_X)$
 if there are chain maps $i: (Y,\partial_Y) \rightarrow (X,\partial_X)$ and $p: (X,\partial_X) \rightarrow (Y,\partial_Y)$, such that $p\circ i = 1$ and
\begin{equation}\label{HomotopyRel}
1 - i\circ p = \{\partial,h\} = \partial \circ h + h \circ  \partial  \, . 
\end{equation} 
A direct  computation with $p$ and $i$ defined in \eqref{Rpro} and \eqref{Rinc} shows  that
\begin{equation}
\partial h(f) = f \, , \qquad h\partial(\phi) = \phi - i  p(\phi) \, . 
\end{equation}
More specifically, the first relation is a direct consequence of the property (\ref{Kernel}), 
while the second relation requires two integrations by part in order to use that the Green's  function 
obeys  the analogue of (\ref{Kernel}) with respect to its second argument. In this computation one picks up boundary terms that 
precisely constitute $- i  p(\phi)$. We thus obtain the homotopy relation \eqref{HomotopyRel} if we 
define   $p$ and $i$ to be zero on $V^{1}$.
This shows that the maps from $V^{\bullet}$ to $\mathbb{R}^2$  define a homotopy retract.

We should emphasize  that the above choice of Green's function and homotopy is just one of many. 
The kernel $K$ of the above  retarded Green's function is, in particular, 
not symmetric under $t \leftrightarrow s$. We can also choose Dirichlet boundary conditions at $t_i$ and $t_f$, for which the 
kernel $K_{DD}(t,s)$ is symmetric and given by 
\begin{equation}\label{KernelKDDD}
K_{DD}(t,s) = \theta(t - s)\frac{\sin \omega(t-s)}{\omega} - \frac{\sin \omega (t - t_i)}{\sin \omega(t_f - t_i)}\frac{\sin \omega (t_f - s)}{\omega} \, .
\end{equation}
This yields the homotopy map 
\begin{equation}\label{DirichletHomotopy}
h(f) = \int_{t_i}^t \text d s \, f(s) \frac{\sin \omega(t-s)}{\omega} - \frac{\sin \omega (t-t_i)}{\sin \omega(t_f - t_i)} \int_{t_i}^{t_f} \text d s \, f(s) \frac{\sin \omega(t_f - s)}{\omega} \,. 
\end{equation}
Although in the 
form (\ref{KernelKDDD}) the symmetry is not manifest, a non-trivial computation
 shows $K_{DD}(t,s) = K_{DD}(s,t)$.\footnote{To this end one uses 
that the step function obeys $\theta(t-s)=1-\theta(s-t)$ in order to create the first term with $s$ and $t$ interchanged. 
The remaining terms can then be rewritten by use of the identity 
 \be
  \sin(x-y) = \sin x \,\cos y - \cos x\, \sin y\;, 
 \ee
after which all terms can be recombined to yield   $K_{DD}(s,t)$. }
The corresponding chain maps are
\begin{equation}
p(\phi) = (\phi(t_i),\phi(t_f))\,, 
\end{equation}
and
\begin{equation}\label{Dinclusion}
i(x_i,x_f) = x_i \frac{\sin \omega (t_f - t)}{\sin \omega(t_f - t_i)} + x_f \frac{\sin \omega (t-t_i)}{\sin \omega(t_f - t_i)} \, .
\end{equation}

As a next step we lift the homotopy retract from the underlying spaces $V^{\bullet}$ and $\mathbb{R}^2$ 
to the BV complex, i.e., to  the space of functionals on $V^{\bullet}$ and the space of functions on $\mathbb{R}^2$. 
We recall that the differential on this complex is given by the BV differential, 
which reads for the harmonic oscillator 
\begin{equation}
Q = \int_{t_i}^{t_f} \text d t ( \ddot \phi(t) + \omega^2 \phi(t)) \frac{\delta}{\delta \phi^*(t)}\, .
\end{equation}
In order to realize the homotopy retract on the BV complex we have to use that, 
as explained above, the maps $i$ and $p$ can be applied to functionals via pullback to give 
$I:=p^*$ and $P:=i^*$. Specifically, 
 for a functional 
$F[\phi, \phi^*]$ in  $\mathcal{F}(V^\bullet)$ we obtain the function on $\mathbb{R}^2$: 
\begin{equation}
P(F)(q,p):= i^{*}(F)(q,p) = F[i(q,p),0] \, , 
\end{equation}
while for a function $f \in \mathcal{F}(\mathbb{R}^2)$ we obtain the functional  on $V^\bullet $ 
\begin{equation}
I(f)[\phi, \phi^*] : = p^{*}(f)[\phi,\phi^*] = f(p(\phi)) = f(\phi(t_i),\phi^*(t_i)) \, .
\end{equation} 
We next have to show how to 
use the homotopy $h$ on $V^\bullet$ to define a homotopy on $\mathcal{F}(V)$. 

We first consider the linear functionals $\phi(t)$ and $\phi^*(t)$. 
To clarify this notation let us point out that the symbol `$\phi(t)$' can be interpreted in two natural ways. 
The standard interpretation is that of a function $\phi$ that takes different values depending on  
the variable $t$: here we think of $\phi$ as fixed and of $t$ as a variable. 
The second interpretation of `$\phi(t)$' takes $t$ to be fixed but $\phi$ to be variable. This then defines a \textit{functional}: 
a map that assigns a number to each function $\phi$, namely the number $\phi(t)$ obtained by evaluating the function 
$\phi$ at the fixed $t$. Similar remarks apply to the interpretation of `$\phi^*(t)$' as a function or as a functional. 
A subtlety of this notation is that the degree of $\phi^*(t)$ depends on its interpretation: as a function it has 
degree $+1$, but as a functional it has degree $-1$, c.f.~the discussion after (\ref{FunctionalMechanics}) above. 
It will always be clear from the context which interpretation applies. 

Returning to the problem of defining a homotopy map $H$ on functionals 
we first define $H$  
on the linear functionals $\phi(t)$ and $\phi^*(t)$ by
\begin{equation}\label{RetardedHomotopy}
\begin{split}
H(\phi(t)) &:= \int_{t_i}^{t} \text d s \,  \frac{\sin \omega(t-s)}{\omega}\phi^*(s) 
\equiv  \int_{t_i}^{t_f} \text d s \,  K(t,s)\, \phi^*(s) \, , \\
H(\phi^*(t)) &:=0\;. 
\end{split}
\end{equation}
Note that $H(\phi(t))$ does not actually depend on $\phi(t)$. 
The functional $H(\phi(t))$  maps a given anti-field $\phi^*$ to the number
that is given by the integral on the right-hand side. This functional is  linear in $\phi^*$ and so 
has degree $-1$, as it should be since  $H$ has intrinsic degree $-1$. 
It is now straightforward to show that
\begin{equation}\label{linearhomotopy}
\begin{split}
QH(\phi(t)) &= \phi(t) - p^*i^* \phi(t)\;, \\
HQ(\phi^*(t)) &= \phi^*(t) \, .
\end{split}
\end{equation}
To obtain a homotopy on all functionals, we observe that any polynomial functional is a superposition of products of the functionals $\phi(t)$ and $\phi^*(t)$. So it is enough to know how the homotopy distributes over any product of functionals. We set
\begin{equation}\label{HomotopyExtension}
H(FG) = \frac{1}{2}\big\{H(F)G + (-)^F p^*i^* (F) H(G) + (-)^{FG} H(G)F + (-)^G p^*i^* (G) H(F)\big\} \, .
\end{equation}
Note that we did a graded symmetrization over $F$ and $G$ on the right-hand side, which is necessary since $H(FG)$ should be graded symmetric in $F$ and $G$. The definition of $H$ ensures that the homotopy relation
\begin{equation}
QH + HQ = 1 - p^*i^* \equiv 1-IP 
\end{equation}
is satisfied on products $FG$, as long as it holds on $F$ and $G$ individually. The action of $H$ on any functional $F$ can then by successively reduced to its action on $\phi(t)$ and $\phi^*(t)$, where we define it via \eqref{RetardedHomotopy}. This shows that $p^* i^*$ is homotopic to the identity. So $\mathcal{F}(V^\bullet)$ is homotopic to $\mathcal{F}(\mathbb{R}^2)$. We therefore proved that the space of functionals on $V^{\bullet}$ is quasi-isomorphic to the space of functions on phase space $\mathbb{R}^2$.

\section{Comparison with Standard Formulations}

In this section we verify from various angles the main statement (\ref{THESTATEMENT}) that relates the cohomology of the 
BV differential to quantum expectation values. First, we explain the perturbation lemma and show that  it relates 
to Wick contractions in the familiar formulation of quantum mechanics. Second, we give a heuristic argument based 
on the path integral formulation.

\subsection{Perturbation Lemma}  
The homological perturbation lemma provides a recipe to compute the cohomology of the BV differential
and in particular the representative $F'$ appearing in the main statement around (\ref{THESTATEMENT}). 
The perturbation lemma states   the following. Suppose that we have homotopic complexes $(X,\text d_X)$ and $(Y, \text d_Y)$. This means that there are  quasi-isomorphisms $p: (X,\text d_X) \rightarrow (Y, \text d_Y)$ and $i: (Y,\text d_Y) \rightarrow (X, \text d_X)$, together with a homotopy $h: (X,\text d_X) \rightarrow (X,\text d_X)$, such that
\begin{equation}
\text d_X \circ h + h\circ  \text d_X = 1 - i\circ p \, .
\end{equation}
Now assume that we perturb the differential of $X$ by some $\eta$, such that we still have a chain complex, i.e.~$\text d_X' := \text d_X + \eta$ obeys $(\text d_X' )^2=0$. Then there is a differential $\text d_Y'$ on $Y$, so that $(Y,\text d_Y')$ is still homotopic to $(X,\text d_X')$. This means that there are new quasi-isomorphisms $p': (X,\text d_X') \rightarrow (Y, \text d_Y')$ and $i': (Y,\text d'_Y) \rightarrow (X, \text d_X')$ with homotopy $h': (X,\text d_X') \rightarrow (X,\text d_X')$ satisfying the homotopy relation
\begin{equation}
\text d_X' \circ h' + h' \circ \text d_X ' = 1 - i'\circ p' \, .
\end{equation}
The perturbation lemma provides explicit formulas for the perturbed data: 
\begin{equation}\label{PerturbedData}
\begin{split}
p' &= p \circ \sum_{n \ge 0} (-\eta h)^n\, , \qquad \;\;\; i' = \sum_{n \ge 0} (-h \eta)^n \circ i \, , \\
h' &= h \circ \sum_{n \ge 0} (-\eta h)^n \, , \qquad\quad 
 \text d_Y' = \text d_Y + p \circ \sum_{n \ge 1} (-\eta h)^n \circ \eta \circ i \, .
\end{split}
\end{equation}
Here the perturbed differential  $\text d_Y'$ is such that the perturbed projection and inclusion are chain maps, 
which means that 
  \be
    \text d_Y' p' = p' \text d_X'\;, \qquad  \text d_X' i' = i'  \text d_Y'\;.
  \ee 

If the complexes have additional structure, we want the perturbations to preserve these. 
For instance, the space of functionals forms an algebra under multiplication, and the cohomological vector field $Q = \{S,-\}$ acts as a derivation. In order to preserve  these properties we first have to assume that the original homotopy data preserves these. So $i$ and $p$ have to be algebra morphisms, i.e.~$i(FG)=i(F)i(G)$ and $p(FG)=p(F)p(G)$.  
In addition, we have to assume that $h$ is a \emph{strong deformation retract}, which means that $h$ is subject to the side conditions
\begin{equation}\label{SideConditions}
p\circ i = 1 \,, \quad h^2 = 0 \, , \quad p \circ h = 0 \, , \quad h \circ i = 0\, . 
\end{equation}
If these conditions are met, the perturbed chain maps $p'$ and $i'$ will again be algebra morphisms, and $\text d_Y'$ will be a derivation.

We may first apply the homological perturbation lemma to a classical mechanical system. Given an action $S$, we 
split it as $S_0 + S_I$, where $S_0$ is a free theory (quadratic in fields) and $S_I$ is the interaction term. We now think of $(\mathcal{F}(V),Q_0 = \{S_0,-\})$ as our initial complex, which is based on the free action (a sum of  harmonic oscillators).  
We saw that a Green's function defines a homotopy on the complex $(V^\bullet,\partial)$ to phase space, 
satisfying $\{\partial,h\} = 1 - i \circ p$, which in turn defines a homotopy $H$
from $\mathcal{F}(V)$ to $\mathcal{F}(\mathbb{R}^2)$, c.f.~(\ref{HomotopyExtension}), satisfying 
\begin{equation}
Q H + H Q = 1 - I  P \, , \quad I = p^* \, , \quad P = i^*\,. 
\end{equation}
Such a homotopy in terms of a Green's function defines a strong deformation retract, i.e.~the analogue of (\ref{SideConditions}) holds for $H$. 
We then take  the interaction 
term $Q_I = \{S_I,-\}$ as the perturbation. Since the functions on solution space are concentrated on degree zero, there will be no induced differential, but the space $(\mathcal{F}(\mathbb{R}^2),0)$ is still  a trivial complex. Since  $Q_0 + Q_I$ encodes the full interacting equations of motion  $P'$ projects functionals to solutions of the full equations of motion. So this is the role 
of the projector $P'$: it  perturbatively constructs solutions of the interacting theory using the Green's function and a given 
solution of the free theory.

We now turn to our core application for the quantum case. 
The perturbation lemma allows us to show that in perturbation theory any functional $F$ of degree zero has a representative $F'$ in the cohomology of $\delta = \{S,-\} - i\hbar \Delta$ that can be written as $F' = f \circ p = I(f)$ for some function $f$ on phase space.  We use a Green's function to define a homotopy and view $\eta = \{S_I,-\} - i \hbar \Delta$ as a perturbation. 
With the perturbation formulas (\ref{PerturbedData})  we obtain the  new chain maps
\begin{equation}
P' = P \sum_{n \ge 0} (-\eta H)^n \, , \qquad I' = I \, , 
\end{equation}
where the second equation is true  by degree reasons.\footnote{Note that $I$ being unperturbed implies that, in the dual picture,  
$p$ is unperturbed, which means that $p$ does not depend on the interactions.} 
We then define 
 \be\label{F'pert}
   F' := I' P'(F)\;, \qquad  f := P'(F)\;.
 \ee  
The claim is that these are the desired functional and function.  
 Indeed, $F'$ is in the same cohomology class as $F$, since
\begin{equation}
F - F' = (1 - I'P')(F) = \delta (H'(F)) \, , 
\end{equation}
where we used the homotopy relation together with  $H(\delta(F))=0$ for degree reasons. 
So the objects  required by  the statement around (\ref{THESTATEMENT}) are given by 
$F' = I(f) = f \circ p$ and $G=-H'(F)$.

\subsection{Wick Contractions and Perturbations}

The perturbation lemma discussed  above  gives a complete perturbative solution to the 
problem of determining the functional $F'$ that in cohomology  is equivalent to $F$, together with the function $f$. 
However, the required action of the homotopy map $H$ given in  \eqref{HomotopyExtension} 
 is inconvenient for  computations due to the explicit symmetrization. 
The action of the homotopy can be brought to a more manageable form by splitting the field space into 
off-shell and on-shell fields. This has the additional advantage of relating the perturbation lemma 
to  standard computations in quantum field theory using Wick contractions. 
More precisely, for the special case that the homotopy is given by the Feynman propagator, the perturbation lemma 
amounts to Wick's theorem, but the perturbation lemma is more general, as we will see in the next section.

We begin by decomposing the field space by using the projection $p$ and inclusion $i$: 
 \be
  V = ip(V) \oplus (1-ip)(V) =: V_p \oplus V_u\;, 
 \ee 
where $V_p$ is the `physical' subspace and $V_u$ is the `unphysical' subspace. 
Any field can now be written as $\phi = \phi_p + \phi_u$, where $\phi_p$ is a solution to the equations of motion. 
Note that  $\phi_u$ depends on the choice of homotopy and  satisfies the same boundary conditions as the Green's function used to construct the homotopy.  In the dual picture of functionals on $V$ we have the induced decomposition 
 \be
  \mathcal{F}(V)  =: \mathcal{F}_p \oplus \mathcal{F}_u\,, 
 \ee
where $\mathcal{F}_p = \mathcal{F}(V_p)$ is the space of functionals depending on $\phi_p$ only. 
More precisely, these are the functionals of the form (\ref{FunctionalMechanics})  where all $\phi$ are $\phi_p$ 
and there are no $\phi^*$. 
Correspondingly, $\mathcal{F}_u$ is the space of functionals  
that contain 
at least one $\phi_u$ or at least one $\phi^*$.

Next we consider the above decomposition at the level of the (functional) derivatives defining the BV algebra. 
Recall  that the space of physical or on-shell fields is isomorphic to a finite-dimensional space 
(in the present case $V_p \cong \mathbb{R}^2$), so that we can use ordinary coordinates 
$(x,y)$ of $\mathbb{R}^2$ to label a  solution $\phi_{p;x,y}$.
This allows us to introduce partial derivatives along solutions: 
\begin{equation}
(\partial_x{F})[\phi,\phi_*]: = \left.\frac{\text d}{\text d \varepsilon}\right|_{\varepsilon = 0} {F}[\phi + \phi_{p;x+ \varepsilon,y},\phi_*] \, , \qquad (\partial_y{F})[\phi]: =  \left.\frac{\text d}{\text d \varepsilon}\right|_{\varepsilon = 0}
 {F}[\phi + \phi_{p;x,y + \varepsilon},\phi_*] \, .
\end{equation}\label{PhysDir}
We also introduce a functional derivative in the direction of $V_u$ via
\begin{equation}\label{UPhysDir}
\int \text d t \, g_u(t)\frac{\delta {F}[\phi,\phi^*]}{\delta \phi_u(t)} := \left.\frac{\text d}{\text d \varepsilon}\right|_{\varepsilon = 0} 
{F}[\phi + \varepsilon g_u,\phi^*] \, ,
\end{equation}
where $g_u \in V_u$. The formula for  $\frac{\delta {F}[\phi,\phi^*]}{\delta \phi_u(t)}$ is the same as for the functional derivative $\frac{\delta  {F}[\phi,\phi^*]}{\delta \phi(t)}$, the only difference being that the space of functions 
entering the integral 
 is restricted. For $\phi^*$ we do not introduce a new notation, since the $\phi^*$ are unaffected by the decomposition. 
 
 We next give a  chain rule relating  $\frac{\delta}{\delta \phi(t)}$ to $\partial_x,\partial_y,\frac{\delta}{\delta \phi_u(t)}$. 
 Indeed, we can represent a functional $F[\phi]$ as a functional $F[\phi_u,x,y]$ (momentarily suppressing the 
 dependence on $\phi^*$)  
 by inverting  the relation $\phi = \phi_u + \phi_p$, 
\begin{equation}
F[\phi] = F[\phi_u[\phi],x[\phi],y[\phi]]\, .
\end{equation}
A straightforward computation along the lines of familiar (finite-dimensional or functional) calculus 
establishes the chain rule
\begin{equation}\label{infiniteChainrule}
\frac{\delta}{\delta \phi(t)} = \int \text d s \, \frac{\delta \phi_u(s)}{\delta \phi(t)} \frac{\delta}{\delta \phi_u(s)} + \frac{\delta x}{\delta \phi(t)}\frac{\partial}{\partial x} + \frac{\delta y}{\delta \phi(t)}\frac{\partial}{\partial y} \, .
\end{equation}
To illustrate this formula, let us look at the projection and inclusion that correspond to the homotopy with Dirichlet boundary conditions, for which 
\begin{equation}
\phi(t) = \phi_u(t) + x \frac{\sin \omega (t_f - t)}{\sin \omega (t_f - t_i)} + y \frac{\sin \omega (t - t_i)}{\sin \omega (t_f - t_i)} \, , 
\end{equation}
with inverse relation 
\begin{equation}
\phi_u(t) = \phi(t) - \phi(t_i) \frac{\sin \omega (t_f - t)}{\sin \omega (t_f - t_i)} - \phi(t_f) \frac{\sin \omega (t - t_i)}{\sin \omega (t_f - t_i)}, \quad x = \phi(t_i), \quad y = \phi(t_f) \, .
\end{equation}
Hence, (\ref{infiniteChainrule}) gives 
\begin{equation}
\begin{split}
\frac{\delta}{\delta \phi(t)} =&\, \frac{\delta}{\delta \phi_u(t)} - \delta(t - t_i) \int \text d s \, \frac{\sin \omega (t_f - s)}{\sin \omega (t_f - t_i)} \frac{\delta}{\delta \phi_u(s)} - \delta(t - t_f) \int \text d s \,\frac{\sin \omega (s - t_i)}{\sin \omega (t_f - t_i)} \frac{\delta}{\delta \phi_u(s)} \\
&\,+ \delta(t - t_i) \frac{\partial}{\partial x} + \delta(t - t_f) \frac{\partial}{\partial y} \, .
\end{split}
\end{equation}

We can now return to our goal of finding a more convenient form of the homotopy lift. 
As a first step we define  
\begin{equation}\label{V_h}
V_h := \int  \text d t \,\text d s \, \phi^*(t) K(t,s) \frac{\delta}{\delta \phi_u(s)} \, , 
\end{equation}
where $K$ is a Green's function for the harmonic oscillator. 
This is a vector field (on  the infinite-dimensional BV manifold) and hence has a simple action as a derivation. 
We will see that, up to a rescaling, this implements the homotopy action on functionals in $\mathcal{F}_u$. 
In fact,  on the linear functionals $\phi = \phi_u + \phi_p$ and $\phi^*(t)$, $V_h$ agrees with the homotopy $H$ 
defined in (\ref{RetardedHomotopy}), since
\begin{equation}\label{vectorishomotopy}
V_h(\phi_u(t) + \phi_p(t)) = V_h(\phi_u(t)) = \int \text d s K(t,s) \phi^*(s) \, , \qquad V_h(\phi^*(t)) = 0 \, . 
\end{equation}
In order to discuss the general case of non-linear functionals 
we recall  the homological vector field $Q_0$ corresponding to the free theory, 
\begin{equation}\label{Q_0}
Q_0 = \int \text d t \,\big(\ddot\phi_u(t) + \omega^2 \phi_u(t)\big)\frac{\delta}{\delta \phi^*(t)} \, ,
\end{equation}
where we used $\phi = \phi_u + \phi_p$ and that $\phi_p$ satisfies the equations of motion and thus drops out. 
The vector fields $V_h$ and $Q_0$ are both odd and hence their Lie bracket is given by the anticommutator, 
which defines a new vector field: 
\begin{equation}
 \{Q_0,V_h\} = \int \text d t \, \Big\{ \phi_u(t)\frac{\delta}{\delta \phi_u(t)}+ \phi^*(t) \frac{\delta}{\delta \phi^*(t)} \Big\} =: N \;. 
\end{equation}
The expression on the right-hand side is found by a computation using  (\ref{V_h}) and (\ref{Q_0}). 
Specifically, this result follows quickly when ignoring boundary terms, but closer inspection shows that 
also the boundary terms vanish.\footnote{This can be easily checked for specific boundary conditions such as Dirichlet, 
but one can also give a general argument valid for arbitrary boundary conditions as follows. We write the Lie bracket as
\begin{equation}\label{NumberOperator}
\{Q_0,V_h\} = N + B \, ,
\end{equation}
where $B$ encodes possible boundary terms. Since $N$ and $\{Q_0,V_h\}$ are vector fields, so is $B$. We will show 
that $B = 0$. Since in (\ref{vectorishomotopy}) we saw that on the linear functionals $\phi = \phi_u + \phi_p$ 
and $\phi^*(t)$, $V_h$ agrees with the homotopy $H$, we have  $\{Q_0,V_h\} = 1 - IP$ on these functionals. 
Since $IP(\phi_u(t)) = IP(\phi^*(t)) = 0$, we find
\begin{equation}\label{linear}
\{Q_0,V_h\}(\phi_u(t)) = \phi_u(t) \, , \quad \{Q_0,V_h\}(\phi^*(t)) = \phi^*(t) \, .
\end{equation}
Comparing this with \eqref{NumberOperator} acting on $\phi_u$ and $\phi^*$, for which 
$N$ acts as the identity, we learn  $B(\phi_u(t)) = 0$ and 
$B(\phi^*(t)) = 0$, i.e., $B$ is zero on these linear functionals. 
General functionals on $\mathcal{F}_u$ are superpositions of functionals of the form
$f(x,y)F[\phi_u,\phi^*]$, 
where $(x,y)$ are coordinates on $V_p$ and $F[\phi_u,\phi^*]$ is at least linear in $\phi_u$ and $\phi^*$. The vector fields $V_h$ and $Q_0$ act directly on $F[\phi_u,\phi^*]$, ignoring the function $f(x,y)$ in front, 
and so does $\{Q_0,V_h\}$. 
Since $B(\phi_u(t)) = B(\phi^*(t)) = 0$ and $B$ acts as a vector field via the Leibniz rule we have 
$B=0$.} 
The resulting vector field has been called $N$, because it  counts the total number of fields $\phi_u$ 
and anti-fields $\phi^*$: 
\begin{equation}
N(\phi_u(t_1)\cdots \phi_u(t_k)\phi^*(t_{k+1})\cdots \phi^*(t_n)) = n \phi_u(t_1)\cdots \phi_u(t_k)\phi^*(t_{k+1})\cdots \phi^*(t_n) \, .
\end{equation}

We claim that on the space ${\cal F}_u$ of functionals that are at least  linear in $\phi_u$ or $\phi^*$, 
on which $N$ is a positive operator, 
the homotopy map is implemented by 
\begin{equation}
H_u := {N}^{-1} V_h \, , 
\end{equation}
where we have identified $N$ with its eigenvalue (which is always positive on ${\cal F}_u$, so $N^{-1}$ is well-defined). 
Indeed, the homotopy relation is then satisfied: 
\begin{equation}
\{Q_0,H_u\} =  {N}^{-1}\{Q_0,V_h\} = {N}^{-1} N = 1 \, , 
\end{equation}
recalling that the subspace ${\cal F}_u$ is projected to $0$ (or, equivalently, 
the homotopy $H_u$ is a strong deformation retract of $\mathcal{F}_u$ to $0$). 
Finally, we can extend $H_u$ to a homotopy $H$ on the total space 
$\mathcal{F}(V) = \mathcal{F}_u \oplus \mathcal{F}_p$ by 
declaring   $H$ to be zero on $\mathcal{F}_p \subseteq \mathcal{F}(V)$. We then  
have $\{Q_0,H\} = 1 - IP$, which defines   a strong deformation retract from $\mathcal{F}(V)$ to $\mathcal{F}_p$.

Having defined, using the above decomposition,   the lift of the homotopy map $H$  we can now 
revisit the application of the perturbation lemma and relate it to Wick contractions. 
In the first step we consider the free harmonic oscillator with differential $Q_0 = \{S_0,-\}$ and 
view $-i\hbar \Delta$ as the perturbation. Under this perturbation there will  neither be a induced differential on $\mathbb{R}^2$, nor a perturbation to the inclusion $I$, while the perturbed projection is
\begin{equation}\label{P1perturbation}
P_1 = P \sum_{n \ge 0} (i\hbar \Delta H)^n \,, 
\end{equation}
where we denote the perturbed projector  by $P_1$ since below we will consider a second perturbation to be 
denoted $P_2$. 
Since $P_1$ is of degree zero it is zero on functionals with at least one anti-field. On functionals of fields only, $\Delta H$ 
acts as
\begin{equation}\label{DeltaHaction} 
 \Delta H = - \int \text d t\, \text d s \, K(t,s)\frac{\delta^2}{\delta\phi(t)\delta \phi_u(s)}\frac{1}{N} \, .
\end{equation}
Recall that the action of $\frac{\delta}{\delta \phi(t)}$ reduces to $\frac{\delta}{\delta \phi_u(t)}$ when it is integrated against a function satisfying the boundary conditions of $V_u$. The kernel $K(t,s)$ is such a function
and so defining 
\begin{equation}
C := \int \text d t\, \text d s \, K(t,s)\frac{\delta^2}{\delta\phi_u(t)\delta \phi_u(s)} \, , 
\end{equation} 
we have with (\ref{DeltaHaction}) the relation 
 \be\label{DeltaH}
  \Delta H = -C\frac{1}{N}\;. 
 \ee 
 
We now consider  a functional $F$ for fixed $t_1, \ldots, t_m$ of the form
\begin{equation}
F[\phi_u,x,y] =\phi_u(t_1)\cdots \phi_u(t_{m})f(x,y) \, ,
\end{equation}
where $f(x,y)$ is any polynomial in $x$ and $y$. Any other functional is a superposition of these functionals, so the effect of $P_1$ on all functionals can be deduced from its effect on $F$ by linearity. Since $P(\phi_u(t)) = 0$, the only non-zero contribution 
to $P_1$ in (\ref{P1perturbation}) comes from the term where $\sum_{n \ge 0} (i\hbar \Delta H)^n$ eliminates  all fields $\phi_u$. This can only happen when $m$ 
is even, i.e., $m = 2k$, for which only the term with $n = k$ contributes. We then find with (\ref{DeltaH}) 
\begin{equation}
\begin{split}
P_1(F) &= f(x,y)(-i\hbar C)\frac{1}{2} \cdots  (-i\hbar C)\frac{1}{2k-2} (-i\hbar C)\frac{1}{2k} \; \phi_u(t_1)\cdots \phi_u(t_{2k}) \\
&= f(x,y) \frac{1}{k!} \left(-\frac{i\hbar}{2} C\right)^k\phi_u(t_1)\cdots \phi_u(t_{2k}) \, .
\end{split} 
\end{equation}
This implies that on arbitrary functionals we have 
\begin{equation}\label{PrimedProjector}
P_1 = P \exp\left(-\frac{i\hbar}{2} C\right)\;.  
\end{equation}
Note that to apply this formula, we do not need the split $V_p\oplus V_u$, since
\begin{equation}
C = \int \text d t\, \text d s \, K(t,s)\frac{\delta^2}{\delta\phi_u(t)\delta \phi_u(s)}  = \int \text d t \,\text d s \, K(t,s)\frac{\delta^2}{\delta\phi(t)\delta \phi(s)} \, ,
\end{equation}
because  $K(t,s)$ is symmetric. (If $K(t,s)$ were not symmetric, it would 
not satisfy the boundary conditions of $V_u$ in the variable $s$.)

The above results allow us to establish the core relation with the familiar techniques  of perturbative quantum field theory: 
The operator $C$ (which is called $2\partial_P$ in lemma 3.4.1 in \cite{CostelloRenormalization}) generates Wick contractions, and $P_1$ implements Wick's theorem upon interpreting 
$P$ as a normal ordering operation. Of course, $P$ depends on the choice of homotopy $H$, and therefore on the choice of propagator. We will see that, when choosing  the Feynman propagator, $P$ can indeed be interpreted as the usual 
prescription to move ``annihilation operators to the right''.

\bigskip 

\noindent 
\textit{\bf Interacting theory:} \\[1ex] 
So far we have applied the perturbation lemma to the free theory with differential $Q_0$, viewing $-i\hbar \Delta$ 
as the perturbation, and related this to familiar Wick contractions. Next we include the non-linear interactions 
of the classical theory, thereby combining the perturbations 
$Q_I = \{S_I,-\}$ and $-i\hbar \Delta$. 
Thus, we apply the perturbation lemma by taking  $Q_I$ to be a perturbation of $Q_0 - i\hbar \Delta$, 
for which one obtains the perturbed map
\begin{equation}\label{PerturbativePI}
P_2 = P_1\circ \sum_{n \ge 0} (-Q_I H)^n \, .
\end{equation}
The core result following from the perturbation lemma is then that the function $f$ computing 
the quantum expectation value for the functional $F$ is given by 
 \be\label{ffromPurtLemma}
  f=P_2(F)\;. 
 \ee

We will now show that this prescription coincides with conventional quantum perturbation theory, 
i.e., that the operator $P_2$ computes expectation values of the full interacting theory. 
This is established if we can show that $P_2$ is in fact equal to $\tilde P_2$ defined by 
\begin{equation}\label{IntegralPI}
 \tilde P_2(F) = \frac{P_1\left(F\exp\left(\tfrac{i}{\hbar}S_I\right)\right)}{Z} \, , 
\end{equation}
with normalization 
\begin{equation}
Z = P_1\exp\left(\tfrac{i}{\hbar}S_I\right) \, , 
\end{equation} 
because this is the familiar method: performing Wick contractions of the operator $F$ weighted by  $\exp(\frac{i}{\hbar}S_I)$ 
with the interacting action $S_I$.  

The proof that (\ref{IntegralPI}) indeed represents $P_2$ follows  \cite{Doubek:2017naz}, which we repeat in our setting. We first show that (\ref{IntegralPI}) is consistent with the relation $P_2 \circ I = 1$ that $P_2$ obeys, i.e., that 
\begin{equation}\label{P2tilde}
\tilde 
P_2 \circ I = 1 \, . 
\end{equation} 
First, using $HI = 0$, we have $\sum_{n \ge 0}(i\hbar \Delta H)^n I(f) = I(f)$ and therefore with (\ref{P1perturbation}) 
\begin{equation}
\begin{split}
P_1\Big(I(f)e^{i\frac{S_I}{\hbar}}\Big) &= P\Big(\sum_{n \ge 0}(i\hbar \Delta H)^n \Big(I(f) e^{i\frac{S_I}{\hbar}}\Big)\Big)
= P\Big(I(f) \sum_{n \ge 0}(i\hbar \Delta H)^n e^{i\frac{S_I}{\hbar}} \Big) \\ 
&= PI(f) P\Big(\sum_{n \ge 0}(i\hbar \Delta H)^n e^{i\frac{S_I}{\hbar}}\Big) = PI(f) Z = f Z \, .
\end{split}
\end{equation}
Here we used that $PI = 1$ and the fact that $P$ is an algebra morphism, i.e.~$P(FG) = P(F)P(G)$.
We next use  that the perturbed data with projection $P_2$ still obey the homotopy relation: 
\begin{equation}
1 - I\circ P_2 = H_2 \circ \delta + \delta \circ H_2 \,,
\end{equation}
where $\delta = Q_0 + Q_I - i \hbar \Delta$ and $H_2 = H \sum_{n \ge 0} (i\hbar\Delta - Q_I)^n$. We apply $\tilde{P}_2$ to both sides and use that $\tilde{P}_2I  = 1$, c.f.~(\ref{P2tilde}).  Then,
\begin{equation}\label{Compare0}
\tilde P_2 - P_2 = \tilde P_2 H_2\delta + \tilde P_2 \delta H_2 \, . 
\end{equation}
Since $\tilde P_2$ is non-zero only in degree zero, we find that $\tilde P_2 H_2\delta = 0$, so \eqref{Compare0} reduces to 
\begin{equation}
\label{Compare}
\tilde P_2 - P_2 =  \tilde P_2 \delta H_2 \, . 
\end{equation}
We now show that $Z \tilde P_2 \delta = 0$, which then proves that $\tilde P_2 = P_2$. 
To this end we decompose the BV differential as $\delta=\delta_0+Q_I$. We assume that $S_I$ does not contain derivatives. We can then write $Q_I=\{S_{I}, \cdot\}$ due to the absence of boundary terms. We compute, for a generic functional $F$, by  use of (\ref{deltaFailure}) 
 \be
 \begin{split}
  \delta_0\big(e^{\frac{i}{\hbar}S_I}F \big)&=  \delta_0\big(e^{\frac{i}{\hbar}S_I}\big)F 
  +e^{\frac{i}{\hbar}S_I}\delta_0F -i\hbar\big\{e^{\frac{i}{\hbar}S_I}, F\big\} \\
  &= e^{\frac{i}{\hbar}S_I}\delta_0F + e^{\frac{i}{\hbar}S_I}\big\{S_I, F\big\} \\
  &= e^{\frac{i}{\hbar}S_I}\delta F\;. 
 \end{split}
 \ee 
Here we used $\delta_0\big(e^{\frac{i}{\hbar}S_I}\big)=0$, which follows because both $S_0$ and $S_I$ contain 
no anti-fields, and $\{e^X, F\}=e^X\{X, F\}$ for any degree-zero object $X$, which follows from (\ref{Compatibility}). 
We then have
\begin{equation}
Z\tilde P_2( \delta F) = P_1\big(\delta F  e^{\frac{i}{\hbar}S_I}\big) = P_1  \delta_0\big(e^{\frac{i}{\hbar}S_I}F \big) = 0 \, ,
\end{equation}
where we used that $P_1 \delta_0 = 0$, i.e.~that $P_1$ is a chain map with respect to $Q_0 - i \hbar \Delta$, 
as implied by the perturbation lemma. This concludes the proof of $P_2 = \tilde P_2$.

We end this part by summarizing what we have found. First of all, we showed that in perturbation theory, any functional $F$ has a representative of the form $F' = f \circ p$. We showed that the perturbative computation of the $Q_0 - i \hbar \Delta$ cohomology does Wick contractions, just like we would expect in a free theory. Also, we showed that for an interacting theory we can think of the homological perturbation lemma computing expectation values with respect to the free theory, but with functionals $F$ weighted by the interacting part $e^{\frac{i}{\hbar}S_I}$. This is the usual way in which 
Feynman diagrams are computed.

\subsection{Path Integral}

In the previous subsection we saw that Wick's theorem arises as a consequence of the perturbation lemma, proving that the latter 
entails in particular   the usual treatment of quantum theories at the perturbative level. Nevertheless, we want to shed more light on the homological approach by comparing it to the path integral derivation.

In the path integral formulation, we formally compute expectation values of functionals by writing
\begin{equation}\label{PathIntegral}
\bra{y;t_f }T(F[\phi])\ket{x;t_i} = \int_{\phi(t_i) = x}^{ \phi(t_f) = y} {\cal D} \phi \, F[\phi]\, e^{\frac{i}{\hbar}S[\phi]}\;.
\end{equation}
The left-hand side represents how this object is computed in the operator language. Here, $\ket{x;t}$ is an eigenstate of the field operator $\hat{\phi}(t)$ with eigenvalue $x$, i.e.~$\hat{\phi}(t)\ket{x;t} = x\ket{x;t}$. Operators and states can be evolved  in time using the unitary time evolution operator, i.e.
\begin{equation}
e^{-\frac{i}{\hbar}Hs}\ket{x;t} = \ket{x;t+s}\,, \quad e^{\frac{i}{\hbar}Hs}\hat{\phi}(t)e^{-\frac{i}{\hbar}Hs} = \hat{\phi}(t+s)\,.
\end{equation}
When working in the operator formalism, we have to include time ordering indicated by the letter $\text{T}$. However, we will often not write time ordering explicitly unless we want to stress its presence. 

The integral on the right-hand side of \eqref{PathIntegral} is thought of as being performed over all paths with $\phi(t_i) = x$ and $\phi(t_f) = y$. We take the perturbative route and write $S = S_0 + S_I$, where $S_0 = \int \frac{1}{2}\dot \phi^2 - \frac{1}{2}\omega^2 \phi^2$. When boundary conditions are fixed, the operator $\partial^2 + \omega^2$ in $S_0$ is invertible, 
and the integral can be  given a meaning in perturbation theory.

One way to proceed  is to pick  reference boundary conditions, e.g.~$\phi(t_i) = \phi(t_f) = 0$. We then write $\phi = \phi_u + \phi_p$, such that $\phi_p$ is the unique classical solution with the generic boundary conditions $\phi_p(t_i) = x, \phi_p(t_f) = y$, while $\phi_u$ satisfies $\phi_u(t_i) = \phi_u(t_f) = 0$. Explicitly, $\phi_p$ is given by $\phi_p = i(x(t_i),y(t_f))$ 
and can be viewed as a constant in the context of the path integral. We then substitute $\phi = \phi_u + \phi_p$  and assume that the integral measure  is invariant under constant shifts 
so that ${\cal D}\phi={\cal D}\phi_u$. One computes 
\begin{equation}
\bra{y;t_f}F[\phi]\ket{x;t_i} = \int_{\phi_u(t_i) = 0}^{ \phi_u(t_f) = 0} {\cal D} \phi_u \, F[\phi_u + \phi_p]\, 
e^{\frac{i}{\hbar}S_I[\phi_u + \phi_p]} e^{\frac{i}{\hbar}\int_{t_i}^{t_f}\frac{1}{2}(\dot \phi_u^2 - \omega \phi_u^2)} e^{i \partial S(\phi_p)} \; ,
\end{equation}
where 
 \be
   \partial S = \frac{1}{2}\phi_p(t_f)\dot{\phi}_p(t_f) - \frac{1}{2}\phi_p(t_i)\dot{\phi}_p(t_i)\;.
 \ee  
Importantly, the boundary action $\partial S$ does not depend on $\phi_u$, which is due to our choice of  
reference boundary conditions $\phi_u(t_i) = \phi_u(t_f) = 0$. 
Thus, the phase $e^{i \partial S(\phi_p)}$ can be scaled out of the path integral: 
 \be\label{IntPI}
  \bra{y;t_f}F[\phi]\ket{x;t_i} = e^{i \partial S(\phi_p)}  \int_{\phi_u(t_i) = 0}^{ \phi_u(t_f) = 0} {\cal D} \phi_u \, F[\phi_u + \phi_p]\, 
e^{\frac{i}{\hbar}S_I[\phi_u + \phi_p]} e^{\frac{i}{\hbar}\int_{t_i}^{t_f}\frac{1}{2}(\dot \phi_u^2 - \omega \phi_u^2)} \; . 
 \ee
Picking  $F = 1$ we have 
\begin{equation}\label{PINormalization}
\braket{y;t_f|x;t_i} = e^{i \partial S(\phi_p)} 
\int_{\phi_u(t_i) = 0}^{\phi_u(t_f) = 0} {\cal D} \phi_u \, e^{\frac{i}{\hbar}S_I[\phi_u + \phi_p]} e^{\frac{i}{\hbar}\int_{t_i}^{t_f}\frac{1}{2}(\dot \phi_u^2 - \omega \phi_u^2)}  \, .
\end{equation}

Our next goal is to define a map $P_1: \mathcal{F}(V^0) \rightarrow \mathcal{F}(\mathbb{R}^2)$ in terms of 
the path integral that equals the map in \eqref{PrimedProjector} that implemented Wick's theorem. 
We claim that this map can be written, for any functional $\tilde F$, as 
\begin{equation}\label{freePI}
P_1(\tilde F) = \int_{\phi_u(t_i) = 0}^{\phi_u(t_f) = 0} {\cal D} \phi_u\, \tilde F[\phi_u + \phi_p] \, e^{\frac{i}{\hbar}\int_{t_i}^{t_f}\frac{1}{2}(\dot \phi_u^2 - \omega \phi_u^2)} \, . 
\end{equation}
Indeed, once we accept that the path integral is computed by doing Wick contractions using the propagator with the boundary conditions specified on the right hand side of \eqref{freePI} it is clear that this 
is equal to the map \eqref{PrimedProjector}. 
Note that $P_1$ is a function on phase space $\mathbb{R}^2$ parametrized by the boundary conditions put on $\phi_p$.

We can finally give a path integral expression for the normalized quantum expectation value of $F$ that agrees with our above 
result from the perturbation lemma. 
In terms of the map (\ref{freePI}) we find with \eqref{IntPI} and \eqref{PINormalization} that 
\begin{equation}
\frac{\bra{y;t_f}F[\phi]\ket{x;t_i}}{\braket{y;t_f|x;t_i}} = \frac{P_1(F e^{\frac{i}{\hbar}S_I})}{P_1(e^{\frac{i}{\hbar} S_I})} 
\equiv P_2(F)\, ,
\end{equation}
using that the phase $e^{i \partial S(\phi_p)}$ cancels. 
This is indeed the same as the map \eqref{IntegralPI} defined by means of 
the perturbation lemma. This confirms that the conventional interpretation of the (otherwise ill-defined) path integral 
in perturbation theory agrees with the homological formulation.

\section{Harmonic Oscillator and Perturbations}

In this section we illustrate the homological formulation of quantum mechanics  by 
applying it to the one-dimensional harmonic oscillator. Specifically, we compute two-point functions both with respect 
to positions eigenstates and with respect to coherent states. Using  the familiar formulation of quantum mechanics, 
based on Hilbert spaces and operators, we then verify that the homological approach yields the correct  results. 
We hope to  illustrate in this  that for certain computations, as in the case of position eigenstates,  
the homological approach is more transparent.

\subsection{Homological Computation for Position Eigenstates}

For a free theory like the harmonic oscillator all expectation values can be computed in terms of the two-point function via Wick's theorem, which we also obtained  from  the perturbation lemma. For this reason, we apply the homological formulation  
to the two-point function of the harmonic oscillator. 
The BV differential for the quantum harmonic oscillator reads
\begin{equation}\label{harmonicBVdsiff}
\delta = \int_{t_i}^{t_f} \text d t \,\left[ \big(\ddot \phi(t) + \omega^2 \phi(t)\big) \frac{\delta}{\delta \phi^*(t)} + i \hbar \frac{\delta^2}{\delta \phi^*(t)\delta \phi(t)} \right]\, .
\end{equation}
According to the  general formulation of 
section \ref{SectionStatement}, for a 
given a functional $F$ the time-ordered correlation function can be computed via a function $f$ on $\mathbb{R}^2$ 
that in cohomology is 
equivalent to $F$: 
\begin{equation}\label{homologicalStatement}
f(x,y) = \frac{\bra{y}T(F)\ket{x}}{\braket{y|x}}\;. 
\end{equation}
More precisely,  one  determines  a representative $F'$ of $F$ in the $\delta$ cohomology that can be written as $F' = f \circ p$, where $p: V^\bullet \rightarrow \mathbb{R}^2$ is given by $p(\phi,\phi^*) = (\phi(t_i),\phi(t_f))$. This means that $F'$ should only depend on $\phi(t_i)$ and $\phi(t_f)$.

Since we want to compute the two-point function, we consider the functional that, for fixed $t, s\in \mathbb{R}$, is defined by 
\be\label{2pointFunctional}
  F[\phi, \phi^*] = \phi(t)\phi(s)\; .
\ee 
In order for $F' = f\circ p$ to be in the same cohomology as $F$ there should be a $G$ such that $F - F' = \delta G$, where $G$ has degree minus one. 
The perturbation lemma in the form  
(\ref{ffromPurtLemma}) immediately gives us the function $f$ as follows: 
 \be
  f=P_1(F)\;, \qquad P_1= i^*  \exp\left(-\frac{i\hbar}{2} C\right)\;, 
 \ee
where $C$ is defined in terms of the Green's function (\ref{KernelKDDD}) with Dirichlet boundary conditions: 
\be\label{CDD}
 C = \int \text d t \,\text d s \, K_{DD}(t,s)\frac{\delta^2}{\delta\phi(t)\delta \phi(s)} \, . 
\ee 
Note that here $P_2=P_1$ since we are considering the free theory. We thus have 
 \be\label{ffunction}
  f=i^*\left(1-\frac{i\hbar}{2}C\right)\phi(t)\phi(s)\;, 
 \ee
using that the higher-order  terms in $\hbar$ vanish when acting on the functional (\ref{2pointFunctional}) with two $\phi$. 
For the second term on the right-hand side we need to use (\ref{CDD}) to compute  
 \be
  C\big(\phi(t)\phi(s)\big) = 2\,K_{DD}(t,s)\;, 
 \ee 
for which one uses that $K_{DD}$ is symmetric. 
To evaluate then $f(x,y)$ the first term in (\ref{ffunction}) maps $x, y$ via the inclusion map  (\ref{Dinclusion}) to a solution $\phi_p$
with boundary conditions $\phi_p(t_i) = x$ and $\phi_p(t_f) = y$ and then evaluates the functional on this solution. 
Thus, we have 
\begin{equation}\label{Dirichlet2p}
f(x,y) = \prod_{r = t,s} \left\{\frac{\sin \omega(r - t_i)}{\sin\omega (t_f - t_i)} y + \frac{\sin \omega(t_f - r)}{\sin\omega (t_f - t_i)} x \right\} - i \hbar K_{DD}(t,s) \, .
\end{equation}
Via the dictionary (\ref{homologicalStatement}) this gives the desired two-point function.

\subsection{Homological Computation for Coherent States}

In the previous subsection we used the propagator with Dirichlet   boundary conditions to find a certain representative in 
cohomology of the functional  $F= \phi(t)\phi(s)$. We want to see what happens when we instead use different propagators. 
The standard  propagator in quantum field theory is the Feynman propagator, 
which we will investigate here. 

In $(1+0)$-dimensions the Feynman propagator is given by
\begin{equation}\label{FeynmanProp}
h_F(f)(t) = i \int_{t_i}^t \text d s f(s) \frac{e^{-i\omega(t-s)}}{2\omega}  + i \int_{t}^{t_f} \text d s f(s) \frac{e^{i\omega(t-s)}}{2\omega} =: h_+(f)(t) + h_-(f)(t)\, .
\end{equation}
Note that $h_F(f)$ is complex, even when $f$ is real. It is therefore not sufficient to work with the field space $V$, but rather 
we should work with the complexified field space $V\otimes \mathbb{C}$.
We now want to derive the associated inclusion $i$ and projection $p_F$ 
from the homotopy relation $\{\partial,h_F\} = 1 - i_F\circ p_F$. 
Since (\ref{FeynmanProp})  is a Green's function it satisfies $(\partial^2_t + \omega^2) h_F(f)(t) = f(t)$. On the other hand, on equations of motion $\ddot \phi + \omega^2 \phi$ we find
\begin{equation}
h_+(\ddot \phi + \omega^2 \phi)(t) = i \frac{\dot \phi(t)}{2\omega} - i \frac{\dot \phi(t_i)}{2\omega}e^{-i\omega(t- t_i)} + \frac{\phi(t)}{2} - \frac{\phi(t_i)}{2}e^{-i\omega(t- t_i)} \, ,
\end{equation}
\begin{equation}
h_-(\ddot \phi + \omega^2 \phi)(t) = - i\frac{\dot \phi (t)}{2\omega} + i \frac{\dot \phi(t_f)}{2\omega}e^{i \omega (t- t_f)} + \frac{\phi(t)}{2} - \frac{\phi(t_f)}{2}e^{i\omega(t-t_f)} \, , 
\end{equation}
and so for the sum 
\begin{equation}\begin{split}
h_F(\ddot \phi + \omega^2 \phi)(t) = \phi(t) - \frac{1}{2}\left(\phi(t_i) - \frac{\dot{\phi}(t_i)}{i\omega}\right)e^{-i\omega(t-t_i)}
- \frac{1}{2}\left(\phi(t_f) + \frac{\dot{\phi}(t_f)}{i\omega}\right)e^{i\omega(t-t_f)} .
\end{split}
\end{equation}
We next define new (complex) functionals  $a(t)$ and $a^\dagger(t)$ by  
\begin{equation}
\phi(t) = \sqrt{\frac{\hbar}{2\omega}}(a^\dagger(t) + a(t))\,, \qquad \dot{\phi}(t) = i\sqrt{\frac{\hbar\omega}{2}}(a^\dagger(t) - a(t))\,. 
\end{equation}
These expressions are motivated by the mode expansion of the harmonic oscillator, but we should emphasize that 
here these are just regular functions, not quantum operators. 
In particular, the function $a^{\dagger}$ is just the complex conjugate of the function $a$, with the notation just reminding us of 
the usual raising and lowering operators. 
In terms of these we have  
\begin{equation}
h_F(\ddot \phi + \omega^2 \phi)(t) = \phi(t) - \sqrt{\frac{\hbar}{2\omega}}\left(a(t_i) e^{-i\omega(t-t_i)} + a^\dagger(t_f) e^{i \omega(t-t_f)}\right)\, .
\end{equation} 
This suggests that we define a projector $p_F: V \otimes \mathbb{C} \rightarrow \mathbb{C}^2$ by  
\begin{equation} 
\phi  \mapsto (a(t_i),a^\dagger(t_f))\,, \quad \phi^* \mapsto 0 \;, 
\end{equation}
and the inclusion $i_F: \mathbb{C}^2 \rightarrow V^0 \otimes \mathbb{C}$ by 
\begin{equation}
(x,y) \mapsto \sqrt{\frac{\hbar}{2\omega}}\left(x e^{-i\omega (t-t_i)} + y e^{i\omega (t-t_f)}\right) \, ,
\end{equation}
with zero image in $V^1$. With these definitions we have $p_F \circ i_F = 1$.

For a given a field $\phi$ the projector $p_F$ gives   the complex values $a(t_i)$ and $a^\dagger(t_f)$. Let us compare this with the projector associated to the propagator with Dirichlet boundary conditions, which gives 
$(\phi(t_i),\phi(t_f))$. When relating  to canonical quantization, we found that the correlator computed with the latter prescription 
used the states $\ket{x}$ and $\bra{y}$ satisfying $\phi(t_i)\ket{x} = x \ket{x}$ and $\bra{y}\phi(t_f) = \bra{y}y$. Correspondingly, 
when finding a representative $F'$ of the cohomology of some functional $F$ with $F' = f \circ p_F$, we expect that $f$ computes correlators with the in-state $\ket{z}$ being an eigenstate of $a(t_i)$ and the out-state $\bra{w}$ being an eigenstate of $a^\dagger(t_f)$. This will be confirmed in the following. 

The eigenstates $\ket{z}$ of the annihilation operator $a$ are called coherent states. We use the convention $a\ket{z} = z\ket{z}$ and $\bra{z}a^\dagger = \bra{z}z$. With this convention, $\bra{\bar z}$ is the conjugate of $\ket{z}$, where $\bar{z}$ denotes the complex conjugate of $z$. We can check whether it is reasonable that $p_F$ gives rise to correlators with coherent states by looking again at the two-point function. By either applying the perturbation lemma or going through the same steps as in the previous section, we find that the representative $F'$ of the cohomology of $F = \phi(t)\phi(s)$ satisfying $F' = f \circ p_F$ is given by
\begin{equation}
F' = -i\hbar K_F(t,s) + \frac{\hbar}{2\omega}\big(a(t_i) e^{-i\omega (t-t_i)} + a^\dagger(t_f) e^{i\omega (t-t_f)}\big)
\big(a(t_i) e^{-i\omega (s-t_i)} + a^\dagger(t_f) e^{i\omega (s-t_f)}\big) \, .
\end{equation}
We therefore claim that
\begin{equation}\label{Feyn2p}
f(w,z) = \frac{\bra{w}T(\phi(t)\phi(s))\ket{z}}{\braket{w|z}} \, ,
\end{equation}
where
\begin{equation}
f(w,z) = - i\hbar K_F(t,s) + \frac{\hbar}{2\omega}\big(z e^{-i\omega (t-t_i)} + w e^{i\omega (t-t_f)}\big)\big(z e^{-i\omega (s-t_i)} + w e^{i\omega (s-t_f)}\big) \, .
\end{equation}

It is straightforward to verify equation \eqref{Feyn2p} in the familiar operator language of quantum mechanics, 
as we do now. We first recall that Wick's theorem implies 
\begin{equation}\label{operatorWick}
T(\phi(t)\phi(s)) = - i \hbar K_F(t,s) + N(\phi(t)\phi(s)) \, ,
\end{equation}
where $N$ is the normal ordering operation. We have the operator relation 
\begin{equation}
\hat{\phi}(t) = \sqrt{\frac{\hbar}{2\omega}}\big(a(t_i) e^{-i\omega (t-t_i)} + a^\dagger(t_f) e^{i\omega (t-t_f)}\big)\;, 
\end{equation}
where $a$ and $a^{\dagger}$ are now interpreted as the creation  and annihilation  operators of the harmonic oscillator, 
satisfying the familiar commutation relations. 
Usually the above expression appears in textbooks for $t_i = t_f = 0$ and normal ordering is defined with respect to $a := a(0)$ and $a^\dagger := a^\dagger(0)$. But this is the same as normal ordering $a(t_i)$ and $a^\dagger(t_f)$, since they only differ from $a$ and $a^\dagger$ by phases. We can now compute 
\begin{equation}
\bra{w}N(\phi(t)\phi(s))\ket{z} \, 
\end{equation}
by evaluating $a(t_i)$ at ${z}$ and $a^\dagger(t_f)$ at $w$. This follows because  $N$ moves  all annihilation operators to the right, where we can then use $a(t_i)\ket{z} = {z}\ket{z}$. Similarly, when creation operators are on the right, we can use $\bra{w}a^\dagger(t_f) = \bra{w}w$. Therefore,
\begin{equation}\label{CoherentNormalOrdering}
\bra{w}N(\phi(t)\phi(s))\ket{z} = \frac{\hbar}{2\omega}\big(z e^{-i\omega (t-t_i)} + w e^{i\omega (t-t_f)}\big)
\big(z e^{-i\omega (s-t_i)} + w e^{i\omega (s-t_f)}\big) \braket{z|w} \, . 
\end{equation}
Combining this with \eqref{operatorWick} then proves \eqref{Feyn2p}.

In case of the Feynman propagator, this result explains why the perturbation lemma gives Wick's theorem via the projector $P'$ and $P$ can be interpreted as normal ordering. Recall that $P = i^*$ just evaluates functionals on-shell, with boundary conditions specified by the inclusion map $i$. But this is just what we did in \eqref{CoherentNormalOrdering}. We evaluated $\phi(t)\phi(s)$ on the solution with $a(t) = z$ at $t = t_i$ and $a^\dagger(t) = w$ at $t = t_f$. Of course, there is nothing special about the two-point functions considered here, 
and so the perturbation lemma says that $P'$ is really Wick's theorem squeezed between coherent states.

\subsection{Comparison with Operator Language}

In the previous two sections we applied the homological recipe to compute correlators with respect to position eigenstates and with respect to coherent states. The respective projectors were given by
\begin{equation}
p(\phi,\phi^*) = (\phi(t_i),\phi(t_f)) \, , \qquad p_F(\phi,\phi^*) = (a(t_i),a^\dagger(t_f)) \, .
\end{equation}
Using Wick's theorem, for $p_F$ it was straightforward to see that our approach agrees with the operator language. For $p$, however, it is harder to verify that  $f$ defined via $F' = f \circ p$ actually computes the correlator with respect to 
position eigenstates, although our formal path integral manipulations above suggest that this must be so.
The general claim following from the homological approach is 
\begin{equation}
f(x,y) = \frac{\bra{y;t_f}T(\phi(t)\phi(s))\ket{x;t_i}}{\braket{y;t_f|x;t_i}}  \, ,
\end{equation}
where $f$ is the function whose pullback $p^*(f)$  equals $F=\phi(t)\phi(s)$ in cohomology. 
Our goal in this subsection is to check this statement using the standard formalism of quantum mechanics.
Since the operator computation is quite involved, for simplicity we set  $x = y = 0$. 
Since the perturbation lemma immediately gives the full result  \eqref{Dirichlet2p}
we see  that  the homological approach is  advantageous in this case. 
When $x = y = 0$, \eqref{Dirichlet2p} equals the Green's function $K_{DD}(t,s)$. So we want to establish the identity
\begin{equation}\label{OperatorKDD}
-i\hbar K_{DD}(t,s) = \frac{\bra{y=0;t_f}T(\phi(t)\phi(s))\ket{x=0;t_i}}{\braket{y=0;t_f|x=0;t_i}} =: g(t,s)
\end{equation}
in the operator language.

We first review some more facts about coherent states. As already stated, a coherent state $\ket{z}$ is an eigenstate of the annihilation operator, i.e.
\begin{equation}
\hat{a} \ket{z} = z \ket{z}\ \;\; \text{for all } z \in \mathbb C\, .
\end{equation}
We use the convention that the hermitian conjugate of $\ket{z}$ is $\bra{\bar z}$, so that $\bra{\bar{z}}\hat{a}^\dagger = \bra{\bar{z}} \bar{z}$. 
Given a general state $\ket{\psi}$, its overlap with a coherent state $\bra{z}$ gives a holomorphic function  in $z$,
\begin{equation}
\psi(z) := \braket{z|\psi}.
\end{equation}
The inner product of two such states is
\begin{equation}\label{SBscalar}
\braket{\psi_1|\psi_2} = \frac{1}{\pi} \int \text d^2 z\, \bar{\psi}_1(\bar{z}) \psi_2(z)e^{-|z|^2}\,.
\end{equation}
The Hilbert space equipped with this inner product is called the Segal-Bargmann space. The identity can be written as
\begin{equation}\label{SBIdentity}
1 = \frac{1}{\pi}\int \text d^2 z e^{-|z|^2}\ket{z}\bra{\bar z} \, .
\end{equation}
The creation operator acts by multiplication
since
\begin{equation}
\bra{z}\hat a^\dagger \ket \psi = z \psi(z) \qquad \Rightarrow \qquad (\hat{a}^{\dagger}\psi)(z)=z\psi(z)\,. 
\end{equation}
We can then deduce from  the inner product \eqref{SBscalar} that $\hat a$ acts by differentiation, i.e.
\begin{equation}
\bra{z}\hat{a} \ket \psi = \frac{\partial}{\partial z} \psi(z) \qquad \Rightarrow \qquad (\hat{a}\psi)(z)=\frac{\partial}{\partial z} \psi(z)\,.
\end{equation}
As a consistency check we note that $[\hat{a},\hat{a}^\dagger] = 1$ in this representation. Since the vacuum state $\ket{0}$ 
is annihilated by $\hat{a}$, 
the vacuum is represented by $\psi_0(z)=\braket{z|0}$ that is in fact constant 
(and equal to one if we normalize it). Likewise, the $n$th excited state is
\begin{equation}\label{nexcited}
\braket{z|n} = \frac{1}{\sqrt{n!}}z^n\, , 
\end{equation}
while  a coherent state reads
\begin{equation}
\braket{z|w} = e^{zw}\;. 
\end{equation}

We now come back to the original goal of this section, i.e.~establishing the identity \eqref{OperatorKDD} in the operator language. To do so, we use Wick's theorem
\begin{equation}
T(\phi(t)\phi(s)) = -i\hbar K_F(t,s) + N(\phi(t)\phi(s)) \, ,
\end{equation}
where $K_F$ is the Feynman propagator and $N$ denotes normal ordering.
Using this in \eqref{OperatorKDD}, we find
\begin{equation}\label{KDDKF}
g(t,s)  = -i\hbar K_{F}(t,s) + \frac{\bra{y= 0;t_f}N(\phi(t)\phi(s))\ket{x = 0 ;t_i}}{\braket{y= 0;t_f|x = 0; t_i}} \, .
\end{equation}
In order to compute the overlap involving the normal ordering, we express it in terms of coherent states using \eqref{SBIdentity}. 
We then need to express  $\ket{x = 0}$ in terms of coherent states. For an arbitrary state $\ket{\psi}$ we have
\begin{equation}\label{segalbargmann1}
\braket{x|\psi} = \frac{1}{\pi}\int \text d^2 z\, e^{-\bar z z} \braket{x|z} \braket{\bar z|\psi}\,.
\end{equation}
This formula can be reduced to an integral over the reals. For example, one can show that  \cite{Hall} 
\begin{equation}\label{segalbargmann2}
\braket{x|\psi} = \psi(x) = C e^{-x^2/2}\int \text d y\, e^{-y^2/2}\braket{x + iy|\psi}\,,
\end{equation}
where $C$ is some constant and $\bra{x + iy}$ is a coherent state, and so 
\begin{equation}
\bra{x} = C e^{-x^2/2}\int \text d y e^{-y^2/2}\bra{ x + iy}\, . 
\end{equation}
In particular,
\begin{equation}\label{x0coherentstate}
\bra{x = 0} = C \int \text d y\, e^{-y^2/2}\bra{iy}\,.
\end{equation}
We now use this in \eqref{KDDKF} to compute $g(t,s)$. We first compute the denominator
\begin{equation}\label{o1}
Z := \braket{y = 0;t_f| x = 0; t_i} = \bra{y = 0}e^{i\frac{H}{\hbar}(t_f - t_i)}\ket{x = 0} \, , 
\end{equation} 
where we used the time evolution operator with respect to the Hamiltonian $H= \hbar \omega(a^\dagger a + \frac{1}{2})$
of the harmonic oscillator.
Thus, using (\ref{x0coherentstate}), we will need the time evolution of a coherent state. 
Defining $T = t_f-t_i$, we need to compute $e^{i\frac{H}{\hbar}T}\ket{z}$, which can be done 
by inserting a complete set of eigenstates $\ket{n}$ of the Hamiltonian  and using  the overlap \eqref{nexcited}. With this, we find $e^{i\frac{H}{\hbar}T}\ket{z} = e^{i\frac{\omega}{2}T}\ket{e^{-i\omega T}z}$. Defining  $\lambda := e^{-i\omega T}$ we thus have $e^{i\frac{H}{\hbar}T}\ket{z} = \lambda^{-\frac{1}{2}}\ket{\lambda z}$. Using this together with \eqref{x0coherentstate} we have: 
\begin{equation}
\begin{split}
Z &= C^2\lambda^{-\frac{1}{2}} \int \text d y_1 \text d y_2 e^{-(y_2^2 + y_1^2)/2}\braket{i y_2|-i\lambda y_1}  \\
&= C^2 \lambda^{-\frac{1}{2}}  \int \text d y_1 \text d y_2 e^{-(y_2^2 + y_1^2)/2 + \lambda y_1 y_2} \\
& = \frac{C_1}{\sqrt{\lambda - \lambda^3}} \, ,
\end{split}
\end{equation}
where we performed the Gaussian integral, and $C_1:= 2\pi C^2 $ is another constant that will cancel in the end. Next we turn to the numerator of \eqref{KDDKF}. Expanding  $\phi(t)$ in terms of ladder operators,
\begin{equation}
\phi(t) = \sqrt{\frac{\hbar}{2 \omega}}(a^\dagger e^{i\omega t} + a e^{-i\omega t}) \, , 
\end{equation} 
and  using  this in \eqref{KDDKF}, we need to compute  expectation values of operators quadratic in $a$ and $a^\dagger$.  
For example, we find that
\begin{equation}
\begin{split}
\bra{y = 0;t_f} a^\dagger a \ket{ x= 0;t_i} &= C^2 \lambda^{-\frac{1}{2}} \int \text d y_1 \text d y_2 \,  \lambda y_1 y_2 \, e^{-(y_2^2 + y_1^2)/2 + \lambda y_1 y_2}\\
&= \frac{C_1 \lambda^{\frac{3}{2}}}{(1-\lambda^2)^{\frac{3}{2}}} = Z \frac{\lambda^2}{1 - \lambda^2} \, .
\end{split}
\end{equation}
Similarly, we have
\begin{align}
 \bra{y = 0; t_f} a a \ket{ x = 0; t_i} &=  -Z \frac{e^{i2\omega t_i}}{1-\lambda^2} \, ,\\
  \bra{y = 0; t_f} a^\dagger a^\dagger \ket{ x = 0; t_i} &= -Z \frac{e^{-i2\omega t_f}}{1-\lambda^2}\, .
\end{align}
Since the operators are normal ordered  we do not need to compute  $\bra{y = 0; t_f} a a^\dagger \ket{ x = 0; t_i}$. We can now 
use the above to compute the normal ordered correlator in \eqref{KDDKF}, for which we find after some algebra
\begin{equation}
\begin{split}
g(t,s) =& -i \hbar K_F(t,s) - \hbar \frac{e^{-i 2 \omega t_f}}{1 - \lambda^2} (2\omega)^{-1} e^{i\omega(t+s)} - \hbar \frac{e^{i 2 \omega t_i}}{1-\lambda^2} (2\omega)^{-1} e^{-i \omega(t+s)} \\
 &+ \hbar \frac{\lambda^2}{1-\lambda^2} (2\omega)^{-1}(e^{i\omega (t-s)} + e^{-i\omega(t-s)})\, . 
\end{split}
\end{equation}
In order to relate this to $K_{DD}$ we rewrite the Feynman propagator,
\begin{equation}
\begin{split}
-iK_F(t,s) &= (2\omega)^{-1}\theta(t-s)e^{-i\omega(t-s)} + (2\omega)^{-1}\theta(s-t)e^{i\omega(t-s)} \\
&= -i K_R(t,s) + (2\omega)^{-1}e^{i\omega (t-s)} \, ,
\end{split}
\end{equation}
where $K_R(t,s) = \theta(t-s)\frac{\sin \omega (t-s)}{2\omega}$ is the retarded propagator. This yields 
\begin{equation}
\begin{split}
g(t,s) =& - i\hbar K_R(t,s) - \hbar\frac{e^{-i2\omega t_f}}{1 - \lambda^2} (2\omega)^{-1} e^{i\omega(t+s)}- \hbar\frac{e^{i2\omega t_i}}{1-\lambda^2} (2\omega)^{-1} e^{-i \omega(t+s)} \\
 & + \hbar\frac{1}{1 - \lambda^2}(2\omega)^{-1}e^{i\omega (t-s)} + \hbar \frac{\lambda^2}{1 - \lambda^2}(2\omega)^{-1}e^{-i\omega(t-s)}  \\
=& -i\hbar K_R(t,s) + i\hbar \frac{\cos \omega(t + s - t_i - t_f) - \cos \omega (t-s+t_f - t_i)}{\sin \omega (t_f - t_i)}  \, ,
\end{split}
\end{equation}
where we reintroduced $t_i$ and $t_f$ through $\lambda = e^{-i\omega (t_f - t_i)}$.
We can now make use of the identity
\begin{equation}
\cos \omega ((t-t_i) + (t_f - s)) - \cos \omega ((t-t_i) - (t_f - s)) = -2\sin \omega (t-t_i) \sin \omega(t_f-s)\;, 
\end{equation}
to arrive at
\begin{equation}
\begin{split}
\frac{\bra{y=0;t_f}T(\phi(t)\phi(s))\ket{x=0;t_i}}{\braket{y=0;t_f|x=0;t_i}}
 &= -i \hbar K_R(t,s) + i\hbar \frac{\sin \omega (t_f-s)}{\omega}\frac{\sin \omega (t-t_i)}{\sin \omega (t_f - t_i)} \\
&= -i \hbar K_{DD}(t,s) \, ,
\end{split}
\end{equation}
which is what we wanted to show.

\subsection{General Boundary Conditions}

In previous subsections we exemplified our approach using two different projectors, which where given by $p_{DD}(\phi) = (\phi(t_i),\phi(t_f))$ and $p_F(\phi)= (a(t_i),a^\dagger(t_f))$. Our computations showed that these determine different types of correlation functions.

We now want to generalize to arbitrary linear boundary conditions. More precisely, we look at boundary conditions of the form
\begin{equation}\label{GeneralBC}
x = a \phi(t_i) + b \frac{\dot \phi(t_i)}{\omega} \, , \qquad y = c \phi(t_f) + d \frac{\dot \phi(t_f)}{\omega} \, ,
\end{equation}
where the numbers $(a,b,c,d,x,y)$ can in general be complex. The numbers $(x,y)$ parametrize solutions to the equations of motion. In this way, we obtain a projector
\begin{equation}\label{generalproj}
\begin{split}
p: C^\infty([t_i,t_f]) \otimes \mathbb{C} &\longrightarrow \mathbb{C}^2 \, , \\
\phi &\longmapsto (a \phi(t_i) + b \frac{\dot \phi(t_i)}{\omega}, c \phi(t_f) + d \frac{\dot \phi(t_f)}{\omega}) \, .
\end{split}
\end{equation}
As usual, we extend $p$ to $V^\bullet \otimes \mathbb C$ by setting $p|_{V^1} = 0$. We recover $p_{DD}$ when $(a,b) = (c,d) = (1,0)$, while $p_F$ is given by $(a,b) = (\bar{c},\bar{d}) = (\sqrt{\frac{\omega}{2\hbar}},i\sqrt{\frac{\omega}{2\hbar}})$.

A solution with boundary conditions \eqref{GeneralBC} is given by
\begin{equation}
\phi_{x,y}(t) = \frac{ya\sin \omega(t - t_i) - yb \cos \omega(t - t_i) + xc\sin\omega(t_f - t) + xd \cos\omega(t_f - t)}{(ad-bc)\cos \omega(t_f - t_i) + (a c + bd) \sin \omega(t_f -t_i)} \, ,
\end{equation}
This solution defines an inclusion 
\begin{equation}
\begin{split}
i:\mathbb{C}^2 &\longrightarrow V^0\otimes \mathbb{C} \, , \\ (x,y) &\longmapsto \phi_{x,y} \, .
\end{split}
\end{equation}
which we extend to $V^\bullet$ via the inclusion $V^0 \hookrightarrow V^\bullet$. To find the homotopy $h$ from the identity to $i\circ p$, we note that the homotopies $h_{DD}$ and $h_F$ satisfy  the boundary conditions \eqref{GeneralBC} with $x = y = 0$. So our ansatz for the homotopy $h(f)$ is the unique solution to $\ddot \phi + \omega^2 \phi = f$ satisfying $p \circ h = 0$. It is given by
\begin{equation}
h(f)(t) = \int_{t_i}^t f(s) K_i(t,s) + \int_{t}^{t_f} K_f(t,s) \, ,
\end{equation}
where
\begin{equation}
K_i(t,s) = K_f(s,t) = \frac{(a \sin \omega (s-t_i) - b \cos \omega(s - t_i))(c\sin\omega (t-t_f) - d \cos \omega(t - t_f))}{ (ad-bc)\omega\cos \omega(t_f - t_i) + (ac + bd)\omega\sin \omega (t_f - t_i)} \, .
\end{equation}
In kernel notation, we have
\begin{equation}
K(t,s) = \theta(t-s)K_i(t,s) + \theta(s-t)K_f(s,t) \, .
\end{equation}
Note that $K(t,s)$ is manifestly symmetric in its arguments. A lengthy computation now shows that
\begin{equation}
\begin{split}
h(\ddot \phi + \omega^2 \phi)(t) =\phi(t) &- \bigg(a \phi(t_i) + b \frac{\dot \phi(t_i)}{\omega}\bigg)\frac{c\sin\omega(t_f - t) + d \cos\omega(t_f - t)}{(ad-bc)\cos \omega(t_f - t_i) + (a c + bd) \sin \omega(t_f -t_i)} \\
&-\bigg(c \phi(t_f) + d \frac{\dot \phi(t_f)}{\omega}\bigg)\frac{a\sin \omega(t - t_i) - b \cos \omega(t - t_i)}{(ad-bc)\cos \omega(t_f - t_i) + (a c + bd) \sin \omega(t_f -t_i)} \, ,
\end{split} 
\end{equation}
as well as $\ddot h(f) + \omega^2 h(f) = f$. Therefore, the homotopy relation $\{\partial,h\} = 1 - i \circ p$ is satisfied.

One application using these more general projectors and homotopies would be the computation of correlators with in- and out states living in different representations of the Hilbert space. For example, one could choose $(a,b) = (1,0)$ and $(c,d) = (0,1)$. In this case, the homotopy satisfies Dirichlet boundary conditions at $t = t_i$ and Neumann boundary conditions at $t = t_f$. The associated representative of the cohomology then uses position eigenstates $\ket{x;t_i}$ as in-states and momentum eigenstates $\bra{p;t_f}$ as out-states.

\subsection{Interacting example: 
correlation functions for $\phi^4$ theory}

In this subsection we illustrate the homological formulation and the perturbation lemma  for 
a simple example of an interacting theory: 
the harmonic oscillator, perturbed by a potential of the form
\begin{equation}
V(\phi) = \int_{t_i}^{t_f}\text dt  \left(\frac{\mu}{3!} \phi^3(t) + \frac{\lambda}{4!} \phi^4(t)\right) \, .
\end{equation}
Our goal is to compute the $k$-point functions to linear order in the coupling constants with respect to the ground state of the free harmonic oscillator for the cases $k = 1,2,4$.

\subsubsection*{One-point function}

We begin by computing the one-point function. We thus consider the functional $F[\phi] = \phi(t)$ for some
fixed time $t$ and use the perturbation lemma, with the BV operator split as $\delta=\delta_0+\delta'$, where 
$\delta_0$ is the classical and free BV operator
\begin{equation}\label{classicalfreedelta} 
\delta_0 = \int_{t_i}^{t_f} \text d t \, (\ddot \phi(t) + \omega^2 \phi(t))\frac{\delta}{\delta \phi^*(t)}\;, 
\end{equation}
and the perturbation is given by
\begin{equation}\label{deltaprime} 
\delta' = -i\hbar \Delta + \int_{t_i}^{t_f} \text d t \, \left(\frac{\mu}{2}\phi^2(t) + \frac{\lambda}{3!}\phi^3(t)\right)\frac{\delta}{\delta \phi^*(t)}  =: \delta_q + \delta_I \, , 
\end{equation}
where $\delta_q=-i\hbar \Delta$ denotes the pure quantum part and $\delta_I$ the classical interactions. 
We recall from (\ref{F'pert})  that the function $f$ computing the quantum expectation values 
is given by 
\begin{equation}
f = P'(F) \, ,
\end{equation}
where
\begin{equation}\label{PerturbationLemmaSum}
P' = P\sum_{n \ge 0} (-\delta' H)^n \, , 
\end{equation}
and $H$ is the homotopy with respect to the classical and free BV operator (\ref{classicalfreedelta}). Recall also that the homotopy is derived from the Green's function of the operator $\partial = \partial_t^2 + \omega^2$ 
on the original complex and satisfies the relation
$h \partial = 1 - ip$, where the projector $p: C^\infty([i_i, t_f]) \otimes \mathbb{C} \rightarrow\mathbb{C}$ is given in \eqref{generalproj}. Explicitly, 
\begin{equation}
p(\phi) = \Big(a\phi(t_i) + b \frac{\dot \phi(t_i)}{\omega}, c \phi(t_f) + d \frac{\phi(t_f)}{\omega}\Big) \ ,
\end{equation} 
where $a,b,c,d$ are complex numbers. On the other hand, the inclusion $i: \mathbb{C}^2 \rightarrow C^\infty([i_i, t_f]) \otimes \mathbb{C}$ maps a pair of boundary values $(a,b)$ to 
the solution $\phi_{a,b} := i(a,b)$ with boundary conditions $p(\phi_{a,b}) = (a,b)$.

Writing  $P' = \sum_{n \ge 0} P_n$ and $f_n = P_n(F)$, to lowest order we have 
\be
 f_0 = P(F) = i^*(F) = F\circ i \;, 
\ee
so that evaluated on $(a,b)$ we find for $F =\phi(t)$ and the inclusion $i$ defined above: 
\begin{equation}
\begin{split}
f_0(a,b) = (F\circ i)(a,b) 
 = \phi_{a,b}(t) \, . 
\end{split} 
\end{equation}

To compute the higher contributions, we recall from (\ref{HomotopyExtension}) 
that
\begin{equation}
H(\phi(t)) := \int_{t_i}^{t_f} \text d u \, K(u,t)\phi^*(u) \, .
\end{equation}
Acting with $\delta'$ defined in (\ref{deltaprime}), we obtain
\begin{equation}\label{deltaHF1}
\begin{split}
-\delta'H(F) = -\int_{t_i}^{t_f} \text d u \, K(u,t)\left(\frac{\mu}{2}\phi^2(u) + \frac{\lambda}{3!}\phi^3(u)\right)\;. 
\end{split}
\end{equation}
We next split 
 \be\label{RemainingSum1}
  P' = P - P \sum_{n\ge 0}(-\delta'H)^n\delta'H\;, 
 \ee
and act on (\ref{deltaHF1})  with $P\sum_{n \ge 0}(-\delta' H)^n$, where to linear order in the coupling constants $\delta' $ can be replaced 
by  $\delta_q$. In this case, we know that \eqref{RemainingSum1} acts via Wick contractions as \eqref{PrimedProjector}. We then find that
\begin{equation}
\begin{split} 
f_1(a,b) &= -\int_{t_i}^{t_f} \text d u \, K(u,t)\left(\frac{\mu}{2}\phi_{a,b}^2(u) + \frac{\lambda}{3!}\phi^3_{a,b}(u)\right)\;, \\
f_2(a,b) &= \int_{t_i}^{t_f} \text d u \, K(u,t)\left(\frac{i\hbar\mu}{2}K(u,u) + \frac{i\hbar\lambda}{2}K(u,u)\phi_{a,b}(u)\right)\;. 
\end{split} 
\end{equation}
All $f_{k \ge 3}(a,b)$ are zero to linear order in the coupling constants.

We learned that the one-point function for general boundary conditions and to linear order in coupling constants is given by
\begin{equation}\label{fullonepoint}
f(a,b) = f_0(a,b) + f_1(a,b) + f_2(a,b) \, .
\end{equation}
To compute the expectation value with respect to the ground state, we should choose $K  = K_F$ to be the Feynman propagator, in which case $a$ and $b$ are the eigenvalues with respect to the annihilation operator, i.e.~they correspond to coherent states. Therefore, if we want the in- and out-state to be both the ground state we 
set $a = b = 0$. In this case $\phi_{a,b} = 0$, and the one-point function reduces to
\begin{equation}
f(0,0) = \frac{\langle 0| \phi(t)|0 \rangle}{\langle 0| 0\rangle}  
= \frac{i\hbar \mu}{2}\int_{t_i}^{t_f} \text d u \, K_F(t,u)K_F(u,u)  \, .
\end{equation}
Note that $K_F(u,u) = \frac{i}{2\omega}$, which in contrast to genuine  quantum field theories is  finite. The above expression gives the tadpole contribution coming from the cubic vertex.

\subsubsection*{Two-point function}

In order  to compute the two-point function we consider the functional $F[\phi] = \phi(s)\phi(t)$ for 
fixed times $s, t$. As for the one-point function, we use the perturbation lemma to compute
\begin{equation}
f(a,b) = P'(F) \, ,
\end{equation}
as a function of the in- and out-states.

We write $P' = \sum_{n \ge 0} P_n$ and $f_n = P_n(F)$, and to lowest order we have 
\begin{equation}
\begin{split}
f_0(a,b) = (F\circ i)(a,b) 
 = \phi_{a,b}(s)\phi_{a,b}(t)\,. 
\end{split} 
\end{equation}
We recall from (\ref{HomotopyExtension}) that
\begin{equation}
H(\phi(s)\phi(t)) := \frac{1}{2}\int_{t_i}^{t_f} \text d u \left( K(u,s)\phi^*(u)(1 + p^* i^*)\phi(t) +  K(u,t)\phi^*(u)(1 + p^*i^*)\phi(s)\right) \, .
\end{equation}
From this we obtain
\begin{equation}\label{deltaHF} 
\begin{split}
-\delta'H(F) = - i\hbar K(t,s) -\int_{t_i}^{t_f} \text d u \, \frac{1}{2}(1 + p^* i^*)\phi(s)K(t,u)\left(\frac{\mu}{2}\phi^2(u) + \frac{\lambda}{3!}\phi^3(u)\right) - (t \leftrightarrow s)\;. 
\end{split}
\end{equation}
We then use the split \eqref{RemainingSum1} and act on (\ref{deltaHF})  with $P\sum_{n \ge 0}(-\delta' H)^n$, which, since \eqref{deltaHF} is already linear in couplings, acts via Wick contractions. We find
\begin{equation}
f_1(a,b) = - i\hbar K(t,s) - \int_{t_i}^{t_f} \text d u \, \phi_{a,b}(s)K(t,u)\left(\frac{\mu}{2}\phi_{a,b}^2(u) + \frac{\lambda}{3!}\phi^3_{a,b}(u)\right) - (t \leftrightarrow s)\,, 
\end{equation}
\begin{equation}
\begin{split}
f_2(a,b) = &\int_{t_i}^{t_f} \text d u \, \phi_{a,b}(s)K(t,u)\left(\frac{i\hbar\mu}{2}K(u,u) + \frac{i\hbar\lambda}{2}\phi_{a,b}(u)K(u,u)\right) - (t \leftrightarrow s) \\
+& \int_{t_i}^{t_f} \text d u \, K(s,u)K(t,u)\left(i\hbar \mu \phi_{a,b}(u) + \frac{i\hbar \lambda}{2}\phi^2_{a,b}(u)\right)\,, 
\end{split}
\end{equation}
\begin{equation}
f_3(a,b) = \int_{t_i}^{t_f} \text d u \frac{\hbar^2 \lambda}{2} K(s,u)K(t,u)K(u,u)\;. 
\end{equation}
All $f_{k \ge 4}(a,b)$ vanish to linear order in the coupling constants. The two-point function for general boundary conditions and to linear order in coupling constants is therefore given by
\begin{equation}\label{fulltwopoint}
f(a,b) = f_0(a,b) + f_1(a,b) + f_2(a,b) + f_3(a,b) \, .
\end{equation}

To compute the two-point function with respect to the ground state we take the Feynman propagator $K = K_F$ and evaluate $f(a,b)$ at $a = b = 0$. Since in this case $\phi_{a,b} = 0$, the two-point function reduces to
\begin{equation}\label{Feynmantwopoint}
\begin{split} 
f(0,0) &= \frac{\langle 0|{\rm T}  \phi(s)\phi(t)|0 \rangle}{\langle 0| 0\rangle}  \\
&= -i\hbar K_F(t,s) + \frac{\hbar^2 \lambda}{2}\int_{t_i}^{t_f} \text d u K_F(s,u)K_F(t,u)K_F(u,u) =: k(s,t) \, .
\end{split} 
\end{equation}

\subsubsection*{Four-point function}
For the four-point function  we need to consider  the functional
\begin{equation}
F_1[\phi]  = \phi(t_1)\phi(t_2)\phi(t_3)\phi(t_4) \, .
\end{equation}
In principle, we could directly use the perturbation lemma as before. However, it turns out to be  easier to first replace $F$ with an equivalent functional $F_2 = \delta G + F_1$ and then apply the perturbation lemma. This illustrates  another strength of the homological approach, since it justifies these types of manipulations.

We choose 
\begin{equation}
G = \int_{t_i}^{t_f} \text d u \ K(t_1,u) \phi^*(u) \phi(t_2)\phi(t_3)\phi(t_4) \, , 
\end{equation}
for which one finds 
\begin{equation}
\begin{split}
\delta G_1 &= i \hbar (K(t_1,t_2)\phi(t_3)\phi(t_4) + K(t_1,t_3)\phi(t_2)\phi(t_4) + K(t_1,t_4)\phi(t_2)\phi(t_3)) \\
& \quad + \int_{t_i}^{t_f} \text d u \ K(t_1,u) \left(\frac{\mu}{2}\phi^2(u) + \frac{\lambda}{3!} \phi^3(u)\right) \phi(t_2)\phi(t_3)\phi(t_4) \\
& \quad + \phi(t_1)\phi(t_2)\phi(t_3)\phi(t_4) - (p^* i^*\phi)(t_1)\phi(t_2)\phi(t_3)\phi(t_4) \, .
\end{split}
\end{equation}
From this we deduce that the homology of $F_1$ is equal to the homology of
\begin{equation}\label{fourpointf2}
\begin{split}
F_2 &= -i \hbar (K(t_1,t_2)\phi(t_3)\phi(t_4) + K(t_1,t_3)\phi(t_2)\phi(t_4) + K(t_1,t_4)\phi(t_2)\phi(t_3)) \\
&\quad - \int_{t_i}^{t_f} \text d u \ K(t_1,u) \left(\frac{\mu}{2}\phi^2(u) + \frac{\lambda}{3!} \phi^3(u)\right) \phi(t_2)\phi(t_3)\phi(t_4) \\
& \quad  + (p^* i^*\phi)(t_1)\phi(t_2)\phi(t_3)\phi(t_4) \, ,
\end{split}
\end{equation}
so we are free to compute the homology of $F_2$. The first line contains two-point functions, so we can use what we already found in this case. Further, the second line is linear in coupling constants so we can compute the expectation value with respect to the free differential, since we only want to keep terms linear in the couplings. We recall 
that $P'$ in \eqref{PerturbationLemmaSum} produces Wick's theorem, so we can immediately compute the expectation value of that part. For the third line, we will do one more manipulation.

Let us write $R_2 = (p^* i^*\phi)(t_1)\phi(t_2)\phi(t_3)\phi(t_4)$ for the term in the third line of \eqref{fourpointf2}. 
Our goal is now again to replace this by the equivalent $R_3 = R_2 + \delta G_2$, where we choose 
\begin{equation}
G_2 = \int_{t_i}^{t_f} \text d u \, K(t_2,u) (p^*i^*\phi)(t_1)\phi^*(u)\phi(t_3)\phi(t_4) \, . 
\end{equation}
This yields 
\begin{equation}
\begin{split}
\delta G_2 &= i\hbar(p^* i^* \phi)(t_1)(K(t_2,t_3)\phi(t_4)+ K(t_2,t_4)\phi(t_3)) \\
& \quad + \int_{t_i}^{t_f} \text d u \, K(t_2,u) (p^*i^*\phi)(t_1)\left(\frac{\mu}{2}\phi^2(u) + \frac{\lambda}{3!}\phi^3(u)\right)\phi(t_3)\phi(t_4) \\
& \quad + (p^* i^*\phi)(t_1)\phi(t_2)\phi(t_3)\phi(t_4) - (p^* i^*\phi)(t_1)(p^* i^*\phi)(t_2)\phi(t_3)\phi(t_4)\;, 
\end{split} 
\end{equation}
from which we obtain 
\begin{equation}\label{4pointnew}
\begin{split}
R_3 &= -i\hbar(p^* i^* \phi)(t_1)(K(t_2,t_3)\phi(t_4)+ K(t_2,t_4)\phi(t_3)) \\
& \quad - \int_{t_i}^{t_f} \text d u \, K(t_2,u) (p^*i^*\phi)(t_1)\left(\frac{\mu}{2}\phi^2(u) + \frac{\lambda}{3!}\phi^3(u)\right)\phi(t_3)\phi(t_4) \\
& \quad + (p^* i^*\phi)(t_1)(p^* i^*\phi)(t_2)\phi(t_3)\phi(t_4) \, . 
\end{split} 
\end{equation}
The second line in \eqref{4pointnew} is again linear in coupling constants, hence we can compute its expectation value with respect to the free theory. The first and third line contain functionals of the general form $(p^*i^* A)B$. For such functionals we will show  that 
\begin{equation}
P'((p^*i^* A)B) = (i^* A) P'(B) \, ,
\end{equation}
where $P'$ is the projector \eqref{PerturbationLemmaSum} computing the expectation value. This follows from the fact that
\begin{equation}
H((p^*i^* A)B) = \frac{1}{2}(Hp^*i^* A)(B + p^*i^* B) + \frac{1}{2}(p^* i^* A + (p^*i^*)^2 A)H(B) = (p^*i^* A) H(B) \, ,
\end{equation}
where we used that $H p^* = 0$ and $i^*p^* = (pi)^* = 1$, together with
\begin{equation}
\delta' ((p^*i^* A) H(B)) = (p^*i^* A) \delta' H(B) \, .
\end{equation}
In the last identity, there is in principle a contribution coming from $i\hbar\Delta$ acting on a field in $(p^*i^* A)$, but these terms vanish due to the boundary conditions of the propagator used in $H$. The upshot is that in the first and last line in \eqref{4pointnew}, we need to know $P'(\phi(t_i))$ and $P'(\phi(t_i)\phi(t_j))$, which are the one- and two-point functions we already computed.

From here on, the rest of the computation can be performed in a straightforward manner. Since the expression is rather lengthy, we will only give parts of it. For example, the free part is given by
\begin{equation}
\begin{split}
P'(\phi(t_1)\phi(t_2)\phi(t_3)\phi(t_4)) &= \phi_{a,b}(t_1)\phi_{a,b}(t_2)\phi_{a,b}(t_3)\phi_{a,b}(t_4) \\
& \quad - i\hbar K(t_1,t_2)\phi_{a,b}(t_3)\phi_{a,b}(t_4) + (t_2 \leftrightarrow t_3) + (t_2 \leftrightarrow t_4) \\
& \quad -i\hbar \phi_{a,b}(t_1)\phi_{a,b}(t_2) K(t_3,t_4) + (t_2 \leftrightarrow t_3) + (t_2 \leftrightarrow t_4) \\
& \quad - \hbar^2 K(t_1,t_2)K(t_3,t_4) + (t_2 \leftrightarrow t_3) + (t_2 \leftrightarrow t_4) + \mathcal{O}(\lambda,\mu)\, .
\end{split}
\end{equation}
The first line is the classical contribution, which is given by the fields at their on-shell value. Comparing to Wick's theorem in the operator language, this is the part where 
no fields are contracted. 
In  the second and third lines there is a total of three contributions each, corresponding to the $s$, $t$ and $u$ channels. 
Again, when comparing to Wick's theorem, this part corresponds to one contraction. 
Finally, the last line corresponds to the fully contracted contribution containing two propagators. It also has a total of three terms corresponding to the different channels. 

When we use the Feynman propagator $K = K_F$ and go to the vacuum by setting $a = b = 0$, only the fully contracted contribution remains. This is also the part one finds using textbook methods. 
For the interacting contributions, we only give the terms that remain after setting $a = b = 0$. In this case, the only terms surviving are
\begin{equation}
\begin{split}
&-i \frac{\hbar^3 \lambda}{2}
K_F(t_3,t_4) \int_{t_i}^{t_f}  \text d u K_F(u,u) K_F(t_1,u)K_F(t_2,u)  + (t_2 \leftrightarrow t_3) + (t_2 \leftrightarrow t_4) \\
&-i \frac{\hbar^3 \lambda}{2}K_F(t_1,t_2)\int_{t_i}^{t_f} \text d u K_F(u,u) K_F(t_3,u)K_F(t_4,u)  + (t_2 \leftrightarrow t_3) + (t_2 \leftrightarrow t_4) \\
&-i\hbar^3 \lambda\int_{t_i}^{t_f} \text d u K_F(t_1,u)K_F(t_2,u)K_F(t_3,u)K_F(t_4,u) \, .
\end{split}
\end{equation}
The first two lines correspond to two disconnected two-point functions, where one of the two-point functions has a one-loop contribution from the $\phi^4$-vertex. There is a term for each channel. The last line is the connected 4-point function, where all four of the $\phi(t_i)$ are connected to the vertex. Combining this with the free contribution, we therefore find that the four-point function in the vacuum and to linear order in the couplings is given by
\begin{equation}
\begin{split}
P'(\phi(t_1)\phi(t_2)\phi(t_3)\phi(t_4)) &=\frac{\langle 0|{\rm T}  \phi(t_1)\phi(t_2)\phi(t_3)\phi(t_4)|0 \rangle}{\langle 0| 0\rangle} \\
&= k(t_1,t_2)k(t_3,t_4) + k(t_1,t_3)k(t_2,t_4) + k(t_1,t_4)k(t_2,t_3) \\
& \quad -i\hbar^3 \lambda\int_{t_i}^{t_f} \text d u K_F(t_1,u)K_F(t_2,u)K_F(t_3,u)K_F(t_4,u) \, ,
\end{split}
\end{equation}
where the second line is written in terms of the two-point functions $k(s,t)$ given in \eqref{Feynmantwopoint}. We obtain the expected result.

\section{Unruh Effect}

In this section we present the first application of the homological formulation 
in the realm of genuine field theories. Specifically, 
we apply our homological method in the context of quantum field theory on curved spacetime 
by  providing an alternative derivation of the Unruh effect: the quantum effect according to 
which the number of particles detected depends on the observer \cite{Unruh:1976db}. In the vacuum state an 
inertial observer in Minkowski space sees no particles, while in the same state  a uniformly accelerated observer sees 
a thermal bath of particles.

\subsection{Generalities and Homotopy Retract}

Let us begin with a brief review of general features of uniformly accelerated observers  in two-dimensional Minkowski spacetime 
with metric 
\begin{equation}
\text ds^2 = \text  dt^2 - \text dx^2  \,. 
\end{equation}
The trajectory of an observer is then parametrized by 
$x^\mu(\tau)= \big( t(\tau), x(\tau)\big)$, where $\tau$ is proper time, so that the 2-velocity $u^\mu(\tau) = dx(\tau)/d\tau$ satisfies the normalization condition
\begin{equation}
\label{2vnormalization}
\eta_{\mu\nu} u^\mu u^\nu = 1 \,. 
\end{equation}  
The Lorentz-invariant condition for the acceleration being constant is expressed in terms of 
 $a^\mu(\tau) = \dot{u}^{\mu}(\tau)$ as 
\begin{equation}
\eta_{\mu\nu} a^\mu (\tau) a^\nu (\tau) = -a^2 \,, 
\end{equation}
where $a$ is a constant. 
The trajectory of a uniformly accelerated observer satisfying 
these two conditions can be written as 
\begin{equation}
t(\tau) = \frac{1}{a} \sinh a \tau \,,\qquad x(\tau) = \frac{1}{a} \cosh a \tau \,. 
\end{equation}

Next, let us relate the inertial frame to a frame that is comoving with the observer. 
This means that denoting these coordinates by $(\tilde{t},\tilde{x})$ the observer's worldline is a 
vertical line $\tilde{x}=0$, so that the observer is indeed at rest in this frame. 
The Rindler coordinates having  this property are defined by 
\begin{flalign}
\label{t in terms of t tilde}
t &= a^{-1} e^{a \tilde{x}} \sinh a \tilde{t} \,,\\
\label{x in terms of x tilde}
x& = a^{-1} e^{a \tilde{x}} \cosh a \tilde{t} \,, 
\end{flalign}
and the inverse relation 
\begin{flalign} 
\tilde{t} &= \frac{1}{2a} \ln \frac{x+t}{x-t} \,,  \\
\tilde{x} & = \frac{1}{2a} \ln \big[ a^2 ( x^2 - t^2 )\big]  \,.
 \end{flalign} 
From these relations one finds  the metric in Rindler coordinates,
\begin{equation}\label{Rindlermetric}
\text ds^2 =(\text  dt)^2 - (\text dx)^2= e^{2a\tilde{x}} \big[ (\text d\tilde{t})^2 - (\text d\tilde{x})^2\big] \,,  
\end{equation}  
which is thus conformally equivalent to the Minkowski metric.

We now consider the action of a massless scalar field $\phi$ in a $1+1$ dimensional spacetime,  
\begin{equation}\label{oscaction}
S [\phi] = \frac{1}{2} \int \text d^2 x \, \sqrt{-g} \, g^{\mu\nu} \partial_\mu\phi  \, \partial_\nu \phi   \,, 
\end{equation}
where $g_{\mu\nu}$ is the metric and $g$ is its determinant. In the inertial frame, 
\begin{equation}
S [\phi] =  \frac{1}{2} \int \text dt\text dx \big[ (\partial_t \phi)^2 - ( \partial_x \phi)^2 \big] \,. 
\end{equation}
The action in the accelerated frame takes the same form: 
\begin{equation}
S [\phi] = \frac{1}{2}\int \text d\tilde{t} \text d\tilde{x} \big[ (\partial_{\tilde{t}} \phi)^2 - ( \partial_{\tilde{x}} \phi )^2 \big]   \,, 
\end{equation}
as a consequence of the conformal invariance of the action (\ref{oscaction}) in two dimensions and the Rindler metric (\ref{Rindlermetric}) being conformally equivalent to the Minkowski metric. The equations of motion are
\begin{equation}
\ddot{\phi}  - \partial_x^2   \phi = 0 \,,
\end{equation}
\begin{equation}
\partial_{\tilde{t}}^2  {\phi}  - \partial_{\tilde{x}}^2  \phi  = 0\,,
\end{equation}
where the dot denotes the partial derivative with respect to time $t$. 
Note that as a scalar we have for the coordinate-transformed field  $\tilde{\phi}(\tilde{t}, \tilde{x})=\phi(t,x)$, so that 
in the second equation we could replace $\phi$ by $\tilde{\phi}$. 

As a preparation for the homotopy retract we have to introduce the Fourier transform  with respect to the spatial coordinate, 
both in inertial and Rindler coordinates: 
 \be
  \phi_k(t) := \int_{-\infty}^{+\infty} \frac{\text dx}{\sqrt{2\pi}} e^{-ikx} \phi(t,x)\;, \qquad
  \tilde{\phi}_l(\tilde{t}) := \int_{-\infty}^{+\infty} \frac{\text d\tilde{x}}{\sqrt{2\pi}} e^{-il\tilde{x}} \tilde{\phi}(\tilde{t},\tilde{x})\;. 
 \ee
Note that even though in the second integral we could replace  $\tilde{\phi}(\tilde{t},\tilde{x})$ by $\phi(t,x)$, 
the Fourier mode  $\phi_k$ as a function of $k$ of course  differs from $\tilde{\phi}_l$ as a function of $l$. 
The inverse relations are 
\begin{equation}
\phi (t,x) = \int_{-\infty}^{+\infty} \frac{\text dk }{\sqrt{2\pi}}e^{ik x}\phi_k  (t)  \;, \qquad 
\tilde{\phi}(\tilde{t},\tilde{x}) =  \int_{-\infty}^{+\infty} \frac{\text  dl}{\sqrt{2\pi}} \,  e^{il\tilde{x} } 
\tilde{\phi}_l (\tilde{t} )\,. 
\end{equation} 
Since the scalar functions on the left-hand sides are  equal (more precisely, we have 
$\phi(t,x)=\tilde{\phi}(\tilde{t}(t,x), \tilde{x}(t,x))$), we have  two different expansions of 
the same $\phi$ into Fourier modes: 
\begin{equation}
\phi (t,x) = \int_{-\infty}^{+\infty} \frac{\text dk }{\sqrt{2\pi}}e^{ik x}\phi_k  (t) \; =  \int_{-\infty}^{+\infty} \frac{\text  dl}{\sqrt{2\pi}} \,  e^{il\tilde{x} (t,x) } 
\tilde{\phi}_l \big(\tilde{t}(t,x) \big)\,. 
\end{equation} 
We will also use the following change of basis for the  Fourier modes and their time derivatives: 
\begin{flalign}
\label{phikttildeaa}
\phi_k  &= \sqrt{\frac{\hbar}{ 2\omega_k }} \big( a^\dagger_{-k}   + a_k\big) \,, \qquad \quad\quad\,\;
\tilde{\phi}_l   = \sqrt{\frac{\hbar}{ 2\Omega_l}} \big( b^\dagger_{-l}   + b_l \big) \,,\\
\label{dotphikktildeaa}
\dot{\phi}_k   & = i \sqrt{\frac{\hbar \omega_k}{2}} \big( a^\dagger_{-k}  - a_{k}  \big)\,, \qquad \quad\;
\partial_{\tilde{t}} \tilde{\phi}_l    = i \sqrt{\frac{\hbar \Omega_l}{2}} \big( b^\dagger_{-l}  - b_{l}  \big) \,, 
\end{flalign}
where  $\omega_k \equiv |k|$, $\Omega_l \equiv |l|$. The inverse relations read:
\begin{flalign}
\label{aakinverserelations}
a_{k}   &= \sqrt{\frac{\omega_{k}}{2\hbar}} \bigg( \phi_k + \frac{i}{\omega_k}{\dot{\phi}_k} \bigg) \,, \qquad 
a_{-k}^\dagger = \sqrt{ \frac{\omega_{k}}{2\hbar}} \bigg( \phi_k -  \frac{i}{\omega_k}{\dot{\phi}_k} \bigg) \,,  \\
\label{bblinverserelations}
b_l  &= \sqrt{\frac{\Omega_l}{2\hbar}} \bigg( \tilde{\phi}_l + \frac{i}{\Omega_l}{\partial_{\tilde{t}} \tilde{\phi}_l} \bigg) \,,  \qquad 
b_{-l}^\dagger = \sqrt{\frac{\Omega_{l}}{2\hbar}} \bigg( \tilde{\phi}_l -  \frac{i}{\Omega_l}{\partial_{\tilde{t}}\tilde{\phi}_l} \bigg) \,. 
\end{flalign}
As for the harmonic oscillator these relations are motivated by the familiar definition of 
creation and annihilation operators, but we emphasize that also here these are just functions.

We now discuss the homotopy retract, beginning with the chain complex defining the theory:
\begin{equation} 
\begin{tikzcd}
0 \arrow[r] & V^0 \arrow[r, "\partial"] & V^{1} \arrow[r] & 0\;. 
\end{tikzcd}
\end{equation}
Here the space of fields  and the space of anti-fields are given by 
\begin{equation}
V^0 = C^\infty ([t_i, t_f] \times \mathbb{R}) \;, \qquad 
V^1 = \Pi C^\infty ([t_i, t_f] \times \mathbb{R}) \,. 
\end{equation}
The notation indicates that the (anti-)fields depend on $t$, restricted to the interval $[t_i, t_f]$, and the 
space coordinate $x$ living on the full real line $\mathbb{R}$.  
The differential is 
\begin{equation}
\partial (\phi) = (\partial_t^2  - \partial_x^2) \phi \,. 
\end{equation} 
The important new feature in field theory is that the 
projector $p \,: V^\bullet \rightarrow C^\infty (\mathbb{R}) \times C^\infty (\mathbb{R})$ no longer  maps 
to a finite-dimensional space like $\mathbb{R}^2$ but to infinite-dimensional functions spaces, however, with 
functions of one less coordinate. Specifically, the projector evaluates  the functions $a$ and $a^\dagger$ 
defined in (\ref{aakinverserelations})
at $t_i$ and $t_f$, respectively: 
\begin{equation}
\label{Unruhprojection} 
\phi  \mapsto \big( a_k(t_i), a_l^\dagger (t_f ) \big)  \,, \qquad \phi^* \mapsto 0 \,. 
\end{equation}
Next, we need to define the inclusion map $i \,: C^\infty (\mathbb{R}) \times C^\infty (\mathbb{R}) \rightarrow V^0$ 
that takes two functions in momentum space, say $c(k)$ and $d(k)$, 
and produces a field in $V^0$ (i.e.~in the present example a scalar 
field in two-dimensional Minkowski space). 
The proper inclusion map satisfying $p\circ i=1$ is given by 
\begin{equation}
\label{aadaggerfginclu}
  \big( c, d \big) \mapsto \phi_{(c,d)}(t,x):= \int_{-\infty}^{+\infty} \frac{dk}{\sqrt{2\pi}} \,  e^{ik {x}}   \sqrt{\frac{\hbar}{ 2\omega_k }} \big( d(-k)  e^{i\omega_k(t - t_f)} +  c(k) e^{-i\omega_k (t-t_i) } \big) \,.
\end{equation} 
The homotopy  map $h: V^1\rightarrow V^0$ is defined, for any $f\in V^1$,
in terms of the Green's function  of the operator $\partial_t^2 - \partial_x^2$: 
\begin{equation}
\label{GreensfunctionforUnruh}
h(f)(t,x) = \int_{t_i}^{t_f } \text ds \int_{-\infty}^{+\infty}  \text dy  \,  K(t-s,x-y)f(s,y)   \,, 
\end{equation} 
where the kernel is explicitly given by 
\begin{equation}
\label{GFk}
 K(t-s, x-y ) = \int_{-\infty}^{+\infty}  \frac{\text dl}{4\pi\omega_l} \, i \big( \Theta (t-s) e^{-i\omega_l (t-s)+ i  l (x-y) } + \Theta(s-t) e^{i\omega_l (t-s) - i l(x-y) } \big) \,. 
\end{equation}
Indeed, one can verify that with the above definitions for projector, inclusion and homotopy 
the homotopy relation $\partial h + h \partial  = 1 - i  p$ is satisfied. 
To this end one needs  to assume that $\phi(t,x)$ and $\partial_x{\phi}(t,x)$ vanish at $x=-\infty$ and $x=+\infty$. 

For completeness we also display the important operations of the dual space of functionals on which 
the BV algebra is defined. 
The BV complex $\mathcal{F}(V^\bullet)$ is equipped with the differential,   
\begin{equation}
Q = \int_{t_i}^{t_f} \text  dt \int_{-\infty}^{\infty} \text dx \,\,\big( \ddot{\phi} ( t,x) + \partial_x^2 \phi ( t,x)\big) \frac{\delta}{\delta \phi^* ( t,x) }  \,. 
\end{equation} 
In addition, the BV-differential is defined as
\begin{equation}
\delta \equiv Q - i \hbar \Delta \,, \qquad \Delta \equiv - \int_{t_i}^{t_f} \text d t\,  
\int_{-\infty}^{\infty} \text d x \,   \,\frac{\delta}{\delta \phi^* ( t, x) } \frac{\delta }{\delta \phi ( t, x) }  \,. 
\end{equation}
For a functional $F[\phi, \phi^*]$ in $\mathcal{F}(V^\bullet)$, we obtain the pull-back 
functional in  $\mathcal{F} [C^\infty ( \mathbb{R}) \times C^\infty ( \mathbb{R})]$ defined by 
\begin{equation}
\label{unruhpullback}
i^*(F) ( c,d) = (F\circ i) (c,d) \,. 
\end{equation}
Similarly, the pullback of a functional $f$ in $\mathcal{F} [C^\infty ( \mathbb{R}) \times C^\infty ( \mathbb{R})]$
with respect to the projection is the functional in $\mathcal{F}(V^\bullet)$ given by 
\begin{equation}
p^* (f) (\phi, \phi^*) = (f \circ p) (\phi, \phi^*)\,. 
\end{equation}

\subsection{Number Expectation Value}

To derive the Unruh effect, one assumes that the number of particles measured by an accelerated observer
is given by the expectation value of the number operator with respect to Rindler space, i.e., 
with respect to creation and annihilation operators defined with the Fourier modes in Rindler space.
More precisely, one computes 
\begin{equation}
{\cal N}_k := \langle N_{k}\rangle \equiv \bra{0} \hat{b}^\dagger_{k} \hat{b}_k \ket{0}\,, 
\end{equation}
where $\hat{b}_k$ and $\hat{b}^\dagger_{k}$ are the Rindler space annihilation and creation operators defined in analogy to 
(\ref{bblinverserelations}), 
and $\ket{0}$ is the Minkowski vacuum state.  This state  is defined so that it is annihilated by the inertial frame 
operator $\hat{a}_k$: 
\begin{equation}
\label{vacuum minkw}
\hat{a}_k \ket{0}  = 0 \,. 
\end{equation} 
For definiteness we take the Heisenberg picture operators $b$ and $b^{\dagger}$ to be at time $\tilde{t}=0$ (which is equivalent to 
$t=0$ for all $x$). 
The usual textbook computation involves relating the creation and annihilation operators of the accelerated and inertial frames  through Bogolyubov transformations. We provide an alternative approach which does not require finding the Bogolyubov transformations. Instead, our strategy is to define the  functional $F[\phi]$ of the massless scalar field $\phi$ to be given by $b_{k}^\dagger b_k$, 
with $b_{k}^\dagger$ and $b_{k}$ being defined in terms of the classical field $\phi$ via (\ref{bblinverserelations}).
Following our approach for the harmonic oscillator in sec. 4, we then find $f(c,d)$ such that $F'=f\circ p$ is in the same cohomology class as $F[\phi]$. Then 
$f(c,d)$ computes  the expectation value
\begin{equation}
f(c,d) = \lim_{\tilde{t} \rightarrow 0}\frac{\bra{d} T \big(\hat{b}_k^\dagger(\tilde{t}) \hat{b}_k (0)\big) \ket{c}}{\braket{d|c}} \;, 
\end{equation}
where $\ket{c}$ and $\ket{d}$ are coherent states with respect to $a_k$, i.e., 
\begin{equation}
a_k \ket{c} = c(k) \ket{c} \,, 
\end{equation} 
and analogously for $\ket{d}$. 
Here we take the limit $\tilde{t}\rightarrow 0$ after performing the computation, as opposed to setting $\tilde{t}=0$ from the beginning, 
since some care is needed in order to deal with  the step functions entering  the Green's function. 
Note that the result does not depend on whether one takes the limit from above or from below, which follows from the 
symmetry of the Green's function. 
Finally, in order to find the expectation value of the Rindler number operator with respect to the Minkowski vacuum, we set $c = d= 0$, 
i.e., 
\begin{equation}
\label{MinkBC}
{\cal N}_k = f(0,0) \,.
\end{equation} 
The choice $c = d= 0$  is the analog of the equation (\ref{vacuum minkw}) specifying the Minkowski vacuum.

We begin by expressing the functional $b^\dagger_k( \tilde{t}) b_k( 0)$ in terms of $\phi(t,x)$. By taking the Fourier transform of (\ref{bblinverserelations}), one obtains $b_k$ and $b_k^\dagger$ in terms of $\phi$ and $\partial_{\tilde{t}}{\phi}$:
\begin{equation}
\label{Rindlerb}
b_k( \tilde{t}) \,\,\,= \int \text  d\tilde{x}\, e^{-ik\tilde{x} } \sqrt{\frac{\Omega_k}{4\pi\hbar}} \bigg( \phi    + \frac{i}{\Omega_k} \partial_{\tilde{t}} {\phi}   \bigg) \,,  
\end{equation}
\begin{equation}
\label{Rindlerbdagger}
b^\dagger_k( \tilde{t}) = \int\text  d\tilde{x}\, e^{ik\tilde{x}}\sqrt{\frac{\Omega_k}{4\pi\hbar}} \bigg( \phi   -  \frac{i}{\Omega_k} \partial_{\tilde{t}} {\phi}   \bigg)   \,.
\end{equation}
For the second equation  we use the chain rule to obtain 
\begin{equation}
\label{phidottilde}
\partial_{\tilde{t}} \phi = \frac{\partial t}{\partial\tilde{t}}\, \dot{\phi }+ \frac{\partial x}{\partial{\tilde{t} }}\, \partial_x \phi  = e^{a\tilde{x}}
\cosh(a\tilde{t}) \dot{\phi} + e^{a\tilde{x}} \sinh(a \tilde{t}) \partial_x \phi  \;. 
\end{equation} 
Note that this is only valid when $x>|t|$ since the Rindler coordinates only cover this part of the Minkowski spacetime. With (\ref{Rindlerb}) -- (\ref{phidottilde}), we can explicitly write out the functional $F[\phi]=b^\dagger_k ( \tilde{t}) b_k ( 0)$:  
\begin{equation}
\begin{split}
\label{Fcorrespondingtobdaggerb}
F[\phi]=
  \int \text d \tilde{x} \int \text d\tilde{y}\, & 
  e^{ik(\tilde{x}-\tilde{y})} \frac{\Omega_k}{4\pi\hbar}\bigg( \phi ( t, x ) \phi ( 0, y )
  + \frac{i}{\Omega_k } e^{a\tilde{y}}   \phi ( t, x)\dot{\phi} ( 0, y )   \\&
 + \frac{1}{\Omega_k^2}e^{a(\tilde{x}+\tilde{y})}  
 \big( \cosh(a \tilde{t})  \dot{\phi} ( 0, y ) \dot{\phi} (t, x) +\sinh( a \tilde{t})  \dot{\phi} ( 0, y ) \partial_x \phi ( t, x) \big)
 \\& 
-\frac{i}{\Omega_k}  e^{a\tilde{x}}\big(  \cosh(a\tilde{t})  \phi ( 0 , y ) \dot{\phi}(t, x) +   \sinh(a \tilde{t})  \phi ( 0 , y ) \partial_x \phi ( t, x)  \big)  \bigg)\, , 
\end{split}
\end{equation} 
where of course $t$ and $x$ on the right-hand side must be viewed as  functions of $(\tilde{t},\tilde{x})$. 

We apply the perturbation lemma to find $f(c,d)$, by using $P_1$ in (\ref{PrimedProjector}),
 \begin{equation} 
P_1 = i^* \exp\left(-\frac{i\hbar}{2} C\right)\;, 
\end{equation} 
where the functional derivatives in the $C$ operator are now with respect to $\phi(t,x)$: 
\begin{equation}
C  = \int\text  d t \, \text dx  \, \text  d s \, \text dy  \,  K(t-s,x-y)\, \frac{\delta^2}{\delta\phi(t,x)\delta \phi(s,y)} \, , 
\end{equation}
and $K(t-s,x-y)$ is given in (\ref{GFk}). 
Applying $P_1$ on $F[\phi]$, 
\begin{equation}
\label{P1F1}
\begin{split}
P_1 (F )(c,d) \, =\,  & \,\,\, i^* F (c,d) \\&-   {i\hbar} \int \text  d \tilde{x} \int\text  d\tilde{y} \,\, e^{ik(\tilde{x}-\tilde{y}) }  \frac{\Omega_k}{4\pi\hbar}
\bigg( K ( t, x -y ) - \frac{i}{\Omega_k } e^{a\tilde{y}} \partial_{t}K ( t, x -y )   \\&
 - \frac{1}{\Omega_k^2}e^{a(\tilde{x}+\tilde{y})}  
 \big[ \cosh (a \tilde{t})\, \partial_{t}\partial_{t} K ( t, x -y )  
+\sinh(a \tilde{t})  \partial_{t} \partial_xK ( t, x -y )\big]
 \\& 
-\frac{i}{\Omega_k}  e^{a\tilde{x}}\big[ \cosh(a \tilde{t})   \partial_{t}K ( t, x -y ) +  \sinh (a \tilde{t})  \partial_x K ( t, x -y ) \big]  \bigg)   \,.
\end{split}
\end{equation}    
There are no further terms in the expansion  of $\exp\left(-\frac{i\hbar}{2} C\right)$ 
because $F[\phi]$ only contains two $\phi$s. Let us start by treating the first term on the right-hand side of (\ref{P1F1}).
Since we set $c=d=0$, the inclusion to the space of fields  (\ref{aadaggerfginclu}) is $i (0, 0  ) =0$. Therefore, with (\ref{unruhpullback}), the first term on the right-hand side of (\ref{P1F1}) vanishes:
\begin{equation}
i^* F(0,0)=0  \,. 
\end{equation}   
Next, we take the limit $t=\tilde{t}=0$. After inserting the derivatives of $K(t,s)$, using (\ref{GFk}),  and writing these in terms of Rindler coordinates, we obtain
\begin{equation}
\begin{split} 
f(0,0)&=P_1(F)(0,0) \\
& =   \int \text d \tilde{x} \int \text d \tilde{y} \int\text  d l \,  \frac{ \Omega_k}{16\pi^2\omega_l}e^{ik(\tilde{x}-\tilde{y})}e^{
 i l a^{-1} ( e^{a\tilde{x}} - e^{a\tilde{y} } ) } 
 \bigg( 1+ \frac{1}{\Omega_k^2} e^{a(\tilde{x}+\tilde{y})}  \omega_l^2      - \frac{\omega_l }{\Omega_k} e^{a\tilde{x}}    
 - \frac{ \omega_l}{\Omega_k} e^{a\tilde{y}}     \bigg) \,. 
\end{split}
\end{equation}
We now  perform the change of variables:
\begin{equation}
u = e^{a\tilde{x}} \,, \quad \frac{1}{au} du  = d \tilde{x} \,, 
\end{equation}
\begin{equation}
v= e^{a\tilde{y}} \,, \quad  \frac{1}{av} du  = d \tilde{y} \,. 
\end{equation}
Then $f(0,0)$ takes the form
\begin{equation}
\begin{split}
f(0,0) = \int_{0}^\infty\text  d u \int_0^\infty \text d v \int_{-\infty}^\infty\text  d l \, e^{ika^{-1} (\ln u - \ln v )  } \frac{ \Omega_k}{16\pi^2 a^2\omega_l}  
e^{  i l a^{-1} ( u  - v) }\bigg( \frac{1}{uv}+ \frac{\omega_l^2    }{\Omega_k^2}    - \frac{\omega_l}{\Omega_k} \frac{1}{v}   - \frac{\omega_l}{\Omega_k} \frac{1}{u}  \bigg)    \,. 
\end{split} 
\end{equation}
Performing  the integrals over $u$ and $v$, recalling $\omega_k \equiv |k|$, $\Omega_l \equiv |l|$,\footnote{
For this computation we used the integral identities:
\begin{equation}
\begin{split}
\int_{0}^{\infty} dx \;e^{iA \ln (x)} e^{iBx} x^{-1} \;  =\;  (-iB)^{-iA} \Gamma(iA)  \,, \qquad 
\int_{0}^{\infty} dx \;e^{iA \ln (x)} e^{iBx}\;  =  \; (iA)(-iB)^{-1-iA} \Gamma(iA)  \,. 
\end{split}
\end{equation}
} yields 
\begin{equation}
\begin{split}
\label{0bb0comp}
&f(0,0)
 =\int_{0}^\infty \text d l \, \,\frac{  \Omega_k}{  4\pi^2 a^2\omega_l}   \Gamma \bigg( - \frac{ik}{a} \bigg)\Gamma \bigg(   \frac{ik}{a} \bigg) (-1)^{ik / a}\,. 
\end{split} 
\end{equation}
As a consistency check, one may verify  that the integrand in (\ref{0bb0comp}) coincides with  the expression 
in equation (8.43) of \cite{Mukhanov:2007zz}. 
By using the following identity for Gamma functions
\begin{equation}
| \Gamma ( ik/a) |^2 =
 \frac{\pi a}{k \sinh (\pi k/a)} = \frac{2\pi a}{|k|} \frac{e^{\pi |k|/a} }{ (e^{2\pi |k|/a}- 1 )} \,, 
\end{equation}
we obtain
\begin{equation} 
\begin{split}
\label{firstboxedequationUNruh}
f(0,0) & =   (e^{2\pi |k|/a}- 1 )^{-1} \int_{0}^\infty \text  dl \,\, \frac{1   }{2\pi a \omega_l}    \,, 
\end{split} 
\end{equation}
as long as we choose $(-1)^{ik/a}=e^{-\pi |k| /a }$.   The expectation value of the number of particles observed by an accelerated observer is a Bose-Einstein distribution with the Unruh temperature
\begin{equation}
T= \frac{\hbar a}{2\pi k_B}\;,  
\end{equation}
where $k_B$ is the Boltzmann constant. 
The divergent integral in (\ref{firstboxedequationUNruh}) is also present in the conventional derivation of the  Unruh effect 
(see, e.g., chapter 8 in \cite{Mukhanov:2007zz}) and is interpreted as the infinite volume of the entire space.

\section{Summary and Outlook}

The main result of this paper is a (partial) reformulation of quantum mechanics that parallels the path integral  
in that there is no reference to Hilbert spaces, states, operators, etc.~However, in contrast to the path integral formulation, 
the homological approach  presented here is algebraic, based on the cohomology of the BV algebra. 
In this one employs a homotopy retract from the infinite-dimensional space of all possible trajectories 
to the finite-dimensional phase space, thereby circumventing the problem to make sense of the path  integral over the full 
infinite-dimensional space. We have shown with a number of examples that the homological formulation allows  one 
to compute concrete quantum expectation values that agree with those determined by standard quantum mechanics. 
However, so far this reformulation is not complete: it only  
provides a prescription  to compute certain normalized quantum expectation values with respect to certain states. 
It then remains as the most important outstanding problem to explore whether this homological formulation could be 
completed to  a full-fledged reformulation of quantum mechanics (and quantum field theory).

It is instructive to compare the techniques  presented here with other approaches in the literature, see 
\cite{Doubek:2017naz,Macrelli:2019afx,Jurco:2019yfd,Arvanitakis:2019ald}.  The idea in these references  
is to pass via homotopy transfer from the $L_{\infty}$-algebra 
of a given theory to a `minimal model' or `on-shell' $L_{\infty}$-algebra  on the cohomology (i.e.~with all differentials being trivial). 
These  $L_{\infty}$ brackets compute (at least tree-level) scattering amplitudes, 
but one has to overcome  some technical challenges. First, in order for the action and inner product to be well-defined 
the space of functions is restricted to Schwartz functions, but then there is no  cohomology, no on-shell fields and hence no 
minimal model (for Schwartz functions  $\square\phi=0$ implies $\phi=0$). One  attempts to circumvent this problem 
by adding on-shell states by hand in degree zero and degree one, so that there is a minimal model. 
However, a priori   the  $L_{\infty}$ brackets are then not well-defined since the product of two on-shell fields 
is neither  on-shell nor a Schwartz function. In order to remedy this ref.~\cite{Macrelli:2019afx} introduces  
certain regularizing factors  in the products of fields. 

In the homological formulation of this paper  these issues do not arise. An important reason is that we do not consider 
smooth functions on $\mathbb{R}$ but rather on the finite integral $[t_i, t_f]$. Then it is not necessary to 
restrict to Schwartz functions and so there is non-trivial cohomology.\footnote{These remarks apply in the realm of 
quantum mechanics where dynamical variables depend only on time. In genuine quantum field theories, the functional dependence 
on the spatial coordinates is probably best chosen to be of  Schwartz type, but  there is still cohomology due to 
the time dependence being more general.} However, since in our formulation the cohomology is concentrated in degree zero 
there is no non-trivial $L_{\infty}$-algebra on this space; rather, the quantum expectation values are computed by 
the functions of the homotopy retract of the BV algebra, as described in the main text.  
Technically, the price to pay for working with general smooth functions on  $[t_i, t_f]$ is that the symplectic form 
encoded in the anti-bracket is no longer invariant under the vector field $Q$. 
However, in our formulation the symplectic form does not enter, and so this does not cause any problems. 
(See also \cite{Cattaneo:2012qu,Cattaneo:2015vsa} where such a more general BV formalism  was developed.)  

We close this paper with a brief list of interesting open problems: 
\begin{itemize}

 \item Most intriguingly, the homological formulation is arguably mathematically well-defined for non-perturbative problems. 
It would then be important to apply and illustrate this method for  genuinely non-perturbative problems. 
 
 \item In this paper we have dealt with theories without gauge symmetries, so it would be interesting to 
 consider gauge field theories such as Yang-Mills theory. Since the BV formalism was originally introduced 
 in order to deal with subtle issues of gauge theories it should be straightforward to set up the corresponding BV algebra. 
 However,  it would still be instructive to work out the homotopy retract and the details of the homological formulation.

 \item One of the potentially most fruitful applications of the homological formulation may arise for 
 quantum field theory on curved spacetime, where traditional flat space techniques to quantization 
 have often  been awkward. Our investigation was in fact motivated by the desire  to find a systematic 
 recipe to obtain in cosmological perturbation theory  the quantum  correlation functions directly 
 from the $L_{\infty}$-algebra or the dual BV algebra. In \cite{Chiaffrino:2020akd} we gave an interpretation 
 of the passing over to gauge invariant so-called Bardeen variables of cosmological perturbation theory 
 in terms of homotopy transfer but 
 it remains to give a similar procedure for the computation of cosmological correlation functions.

\end{itemize}

\subsection*{Conflict of Interest Statement} 

We hereby certify that the research reported in this manuscript and the manuscript itself do 
not provide a conflict of interest for any of the authors.

\subsection*{Acknowledgements}

We would like to thank Roberto Bonezzi, Tomas Codina and Felipe Diaz-Jaramillo for useful discussions 
and Owen Gwilliam and Kasia Rejzner for comments on the first version of this paper. 

This work is funded   by the European Research Council (ERC) under the European Union's Horizon 2020 research and innovation programme (grant agreement No 771862)
and by the Deutsche Forschungsgemeinschaft (DFG, German Research Foundation), ``Rethinking Quantum Field Theory", Projektnummer 417533893/GRK2575.

\appendix

\section{Homological Algebra}

In this work we will frequently use the language of homological algebras. In this appendix we introduce all definitions and facts we will use.

The central object one studies in homological algebra are (co-)chain complexes. A special case are differential graded vector spaces. These are collections of vector spaces $V^n, \, n \in \mathbb{Z}$, together with linear maps $\partial^n : V^n \rightarrow V^{n+1}$ such that $\partial_{n + 1} \circ \partial_n = 0$ for all $n$. This data usually is depicted as a diagram
\begin{equation}
\begin{tikzcd}
\cdots  \arrow[r,"\partial_{n-1}"] & V^n \arrow[r,"\partial_{n}"] & V^{n+1} \arrow[r,"\partial_{n+1}"] & V^{n+2} \arrow[r,"\partial_{n+2}"] & \cdots 
\end{tikzcd} \, .
\end{equation}
If there is an $n$ such that $V^k = 0$ for all $k > n$, we draw it as
\begin{equation}
\begin{tikzcd}
\cdots  \arrow[r,"\partial_{n-2}"] & V^{n-1} \arrow[r,"\partial_{n-1}"] & V^n \arrow[r] & 0
\end{tikzcd} \, ,
\end{equation}
i.e. the sequence ends at $0$ and it is understood that in principle it can be continued by zeros indefinitely to the right. Similarly, if there is an $n$ such that $V^k = 0$ for all $k < n$, we write
\begin{equation}
\begin{tikzcd}
0 \arrow[r] & V^n \arrow[r,"\partial_{n}"] & V^{n+1} \arrow[r,"\partial_{n+1}"] & \cdots 
\end{tikzcd} \, .
\end{equation}
This sequence of vector spaces can also be viewed as the total space $V^\bullet = \bigoplus_{n \in \mathbb{Z}} V^n$, and one then 
defines $\partial: V^\bullet \rightarrow V^\bullet$ via $\partial = \sum_{n \in \mathbb{Z}} \partial_n$. It satisfies $\partial^2 = 0$. To recover the vector subspaces $V_n$, we define a ``charge'' $C: V^\bullet \rightarrow V^\bullet$ via $C(x) = n x$ when $x \in V^n$. The original collection of maps $\partial_n : V^n \rightarrow V^{n+1}$ is then equivalent to a single linear map $\partial: V^\bullet\rightarrow V^\bullet$ 
with operator $C: V^\bullet\rightarrow V^\bullet$ splitting $V^\bullet$ into eigenspaces $V^n$ of integer eigenvalues $n$, and such that $[C,\partial] = 1$. In other words, $\partial$ increases  the charge by one unit. The charge of an element $x \in V^\bullet$ is universally called the degree of $x$.

The fact that $\partial_{n+1} \circ \partial_n = 0$ implies that $\text{im} \, \partial_{n-1} \subseteq \ker \partial_n$. This property allows us to define the \emph{cohomology}
\begin{equation}
H^n(V^\bullet) = \frac{\ker \partial_n}{\text{im} \, \partial_{n-1}} 
\end{equation}
in degree $n$, which are vector spaces by themselves. Equivalently, we can think of $H^\bullet(V^\bullet)$ as a differential graded vector space with
\begin{equation}
\begin{tikzcd}
\cdots \arrow[r,"0"] & H^n(V^\bullet) \arrow[r,"0"] & H^{n+1}(V^\bullet) \arrow[r,"0"] & H^{n+2}(V^\bullet) \arrow[r,"0"] & \cdots 
\end{tikzcd} \, ,
\end{equation}
i.e.~$\partial_n = 0$ for all $n$.

Homomorphisms $f: (V^\bullet,\partial) \rightarrow (W^\bullet,\tilde\partial)$ of chain complexes are collections of linear maps $f_n: V^n \rightarrow W^n$, such that $f_{n+1}\circ \partial_n = \tilde{\partial}_{n}\circ f_n$. This means that the diagram
\begin{equation}
\begin{tikzcd}
\cdots  \arrow[r,"\partial_{n-1}"] & V^n \arrow[r,"\partial_{n}"] \arrow[d,"f_{n}"] & V^{n+1} \arrow[r,"\partial_{n+1}"] \arrow[d,"f_{n+1}"] & V^{n+2} \arrow[r,"\partial_{n+2}"] \arrow[d,"f_{n+2}"] & ...  \\
\arrow[r,"\tilde \partial_{n-1}"] & W^n \arrow[r,"\tilde \partial_{n}"] & W^{n+1} \arrow[r,"\tilde \partial_{n+1}"] & W^{n+2} \arrow[r,"\tilde \partial_{n+2}"] & \cdots 
\end{tikzcd} 
\end{equation}
commutes. In this case $f$ is called a \emph{chain map}. The importance of this definition lies in the fact that $f$ induces a map $H^n(f): H^n(V^\bullet) \rightarrow H^n(V^\bullet)$ on cohomology. Notice that we have $f_n(\ker \partial_{n}) \subseteq \ker \tilde \partial_{n}$, so we can define $\tilde f_n: \ker \partial_n \rightarrow H^n(W^\bullet)$ via $\tilde f_n(x) = f_n(x) \mod \text{im} \, \partial_{n-1}$. On the other hand, we also have $f_n(\text{im} \, \partial_{n-1}) \subseteq \text{im} \, \tilde \partial_{n-1}$, hence $\tilde f_n(\text{im} \, \partial_{n-1}) = 0$. Therefore, $\tilde f_n$ descends to a linear map $H^n(f): H^n(V^\bullet) \rightarrow H^n(W^\bullet)$.

In homological algebra we are mainly interested in the cohomology $H^\bullet(V^\bullet)$ rather than $V^\bullet$ itself. For this reason, we consider differential graded algebras $V^\bullet$ and $W^\bullet$ equivalent, if they have isomorphic cohomologies, i.e.~$H^n(V^\bullet) \cong H^n(W^\bullet)$ for all $n$. We say that $V^\bullet$ and $W^\bullet$ are \emph{quasi-isomorphic}. Along the same line, we say that two chain maps $f,g: V^\bullet \rightarrow W^\bullet$ are quasi-isomorphic, if they agree on homology.

One way to show that linear maps are quasi-isomorphic is to show that they are \emph{homotopic}. We say that chain maps $f,g: (V^\bullet, \partial) \rightarrow (W^\bullet, \tilde \partial)$ are homotopic, if there are maps $h_n: V^n \rightarrow W^{n-1}$, such that
\begin{equation}
f_n - g_n = h_{n+1}\circ \partial_n + \tilde \partial_{n-1}\circ  h_n \, .
\end{equation}
It is straightforward to see that with this condition, $f$ and $g$ agree on cohomology. A related concept are homotopic spaces. We say that $V^\bullet$ and $W^\bullet$ are homotopic, if there are maps $p: (V^\bullet, \partial) \rightarrow (W^\bullet, \tilde \partial)$ and $i: (W^\bullet, \tilde \partial) \rightarrow (V^\bullet, \partial)$, such that $i \circ p$ are homotopic to the identity $\text{id}_{V^\bullet}$ on $V^\bullet$. This then implies that $p$ and $i$ are inverse on homology, hence $V^\bullet$ and $W^\bullet$ are quasi-isomorphic.


\begin{thebibliography}{99}


\bibitem{Batalin:1981jr}
I.~A.~Batalin and G.~A.~Vilkovisky,
``Gauge Algebra and Quantization,''
Phys. Lett. B \textbf{102}, 27-31 (1981)

\bibitem{Batalin:1984jr}
I.~A.~Batalin and G.~A.~Vilkovisky,
``Quantization of Gauge Theories with Linearly Dependent Generators,''
Phys. Rev. D \textbf{28}, 2567-2582 (1983)
[erratum: Phys. Rev. D \textbf{30}, 508 (1984)]



\bibitem{GwilliamThesis} 
Owen Gwilliam, ``Factorization algebras and free field theories," 
PhD thesis, \textit{https://people.math.umass.edu/~gwilliam/thesis.pdf}


\bibitem{GwilliamJF}
O.~Gwilliam and T.~Johnson-Freyd,
``How to derive Feynman diagrams for finite-dimensional integrals
directly from the BV formalism,''
Topology and quantum theory in interaction, 175-185, Contemp. Math., 718, Amer. Math. Soc., Providence, RI, 2018
[arXiv:1202.1554v2 [math-ph]].



\bibitem{CostelloRenormalization}
K.~Costello, ``Renormalization and Effective Field Theory,''
(2011). 

\bibitem{Costello:2016vjw}
K.~Costello and O.~Gwilliam,
``Factorization Algebras in Quantum Field Theory I,'' (2016). 

\bibitem{Gwilliam:2017ses}
O.~Gwilliam and K.~Rejzner,
``Relating Nets and Factorization Algebras of Observables: Free Field Theories,''
Commun. Math. Phys. \textbf{373}, no.1, 107-174 (2020)
[arXiv:1711.06674 [math-ph]].

\bibitem{Brunetti:2013maa}
R.~Brunetti, K.~Fredenhagen and K.~Rejzner,
``Quantum gravity from the point of view of locally covariant quantum field theory,''
Commun. Math. Phys. \textbf{345}, no.3, 741-779 (2016)
[arXiv:1306.1058 [math-ph]].


\bibitem{Okawa:2022sjf}
Y.~Okawa,
``Correlation functions of scalar field theories from homotopy algebras,''
[arXiv:2203.05366 [hep-th]].



\bibitem{Gomis:1994he}
J.~Gomis, J.~Paris and S.~Samuel,
``Antibracket, antifields and gauge theory quantization,''
Phys. Rept. \textbf{259} (1995), 1-145
doi:10.1016/0370-1573(94)00112-G
[arXiv:hep-th/9412228 [hep-th]].




\bibitem{Zwiebach:1992ie}
B.~Zwiebach,
``Closed string field theory: Quantum action and the B-V master equation,''
Nucl. Phys. B \textbf{390}, 33-152 (1993)
doi:10.1016/0550-3213(93)90388-6
[arXiv:hep-th/9206084 [hep-th]].




\bibitem{Lada:1994mn}
    T.~Lada and M.~Markl
    "Strongly homotopy Lie algebras,"
	[hep-th/9406095].

\bibitem{Lada:1992wc}
T.~Lada and J.~Stasheff,
``Introduction to SH Lie algebras for physicists,''
Int. J. Theor. Phys. \textbf{32}, 1087-1104 (1993)
doi:10.1007/BF00671791
[arXiv:hep-th/9209099 [hep-th]].


\bibitem{Alexandrov:1995kv}
M.~Alexandrov, A.~Schwarz, O.~Zaboronsky and M.~Kontsevich,
``The Geometry of the master equation and topological quantum field theory,''
Int. J. Mod. Phys. A \textbf{12}, 1405-1429 (1997)
[arXiv:hep-th/9502010 [hep-th]].

\bibitem{Munster:2011ij}
K.~Munster and I.~Sachs,
``Quantum Open-Closed Homotopy Algebra and String Field Theory,''
Commun. Math. Phys. \textbf{321}, 769-801 (2013)
doi:10.1007/s00220-012-1654-1
[arXiv:1109.4101 [hep-th]].

\bibitem{Hohm:2017pnh}
O.~Hohm and B.~Zwiebach,
``$L_{\infty}$ Algebras and Field Theory,''
Fortsch. Phys. \textbf{65}, no.3-4, 1700014 (2017)
doi:10.1002/prop.201700014
[arXiv:1701.08824 [hep-th]].



\bibitem{Erbin:2020eyc}
H.~Erbin, C.~Maccaferri, M.~Schnabl and J.~Vo\v{s}mera,
``Classical algebraic structures in string theory effective actions,''
JHEP \textbf{11}, 123 (2020)
doi:10.1007/JHEP11(2020)123
[arXiv:2006.16270 [hep-th]].


\bibitem{Koyama:2020qfb}
D.~Koyama, Y.~Okawa and N.~Suzuki,
``Gauge-invariant operators of open bosonic string field theory in the low-energy limit,''
[arXiv:2006.16710 [hep-th]].

\bibitem{Arvanitakis:2020rrk}
A.~S.~Arvanitakis, O.~Hohm, C.~Hull and V.~Lekeu,
``Homotopy Transfer and Effective Field Theory I: Tree-level,''
[arXiv:2007.07942 [hep-th]]; 
``Homotopy Transfer and Effective Field Theory II: Strings and Double Field Theory,''
[arXiv:2106.08343 [hep-th]].







\bibitem{crainic2004perturbation}
  M. ~Crainic,
  ``On the perturbation lemma, and deformations,''
  [arXiv:math/0403266 [math.AT]].



\bibitem{vallette2014algebra}
	B. ~Vallette,
	``Algebra $+$ Homotopy $=$ Operad,''
  Symplectic, Poisson, and noncommutative geometry {\bf 62}, 229 (2014)
	[arXiv:1202.3245 [math.AT]].
	
	
\bibitem{Markl:1997bj}
M.~Markl,
``Loop homotopy algebras in closed string field theory,''
Commun. Math. Phys. \textbf{221}, 367-384 (2001)
doi:10.1007/PL00005575
[arXiv:hep-th/9711045 [hep-th]].	



\bibitem{Cattaneo:2012qu}
A.~S.~Cattaneo, P.~Mnev and N.~Reshetikhin,
``Classical BV theories on manifolds with boundary,''
Commun. Math. Phys. \textbf{332} (2014), 535-603
doi:10.1007/s00220-014-2145-3
[arXiv:1201.0290 [math-ph]].

\bibitem{Cattaneo:2015vsa}
A.~S.~Cattaneo, P.~Mnev and N.~Reshetikhin,
``Perturbative quantum gauge theories on manifolds with boundary,''
Commun. Math. Phys. \textbf{357} (2018) no.2, 631-730
doi:10.1007/s00220-017-3031-6
[arXiv:1507.01221 [math-ph]].


	
\bibitem{Kajiura2003}
H. ~Kajiura, ``Noncommutative homotopy algebras associated with open strings,''
Reviews in Mathematical Physics {\bf 19.01}, 1 (2007)
[arXiv:math/0306332 [math.QA]].

	
	
\bibitem{Doubek:2017naz}
M.~Doubek, B.~Jur\v{c}o and J.~Pulmann,
``Quantum $L_\infty$ Algebras and the Homological Perturbation Lemma,''
Commun. Math. Phys. \textbf{367}, no.1, 215-240 (2019)
doi:10.1007/s00220-019-03375-x
[arXiv:1712.02696 [math-ph]].	

\bibitem{Macrelli:2019afx}
T.~Macrelli, C.~S\"amann and M.~Wolf,
``Scattering amplitude recursion relations in Batalin-Vilkovisky\textendash{}quantizable theories,''
Phys. Rev. D \textbf{100}, no.4, 045017 (2019)
doi:10.1103/PhysRevD.100.045017
[arXiv:1903.05713 [hep-th]].

\bibitem{Jurco:2019yfd}
B.~Jur\v{c}o, T.~Macrelli, C.~S\"amann and M.~Wolf,
``Loop Amplitudes and Quantum Homotopy Algebras,''
JHEP \textbf{07}, 003 (2020)
doi:10.1007/JHEP07(2020)003
[arXiv:1912.06695 [hep-th]].


\bibitem{Arvanitakis:2019ald}
A.~S.~Arvanitakis,
``The L$_\infty$-algebra of the S-matrix,''
JHEP \textbf{07}, 115 (2019)
doi:10.1007/JHEP07(2019)115
[arXiv:1903.05643 [hep-th]].


\bibitem{JF}
T.~Johnson-Freyd,
``Homological perturbation theory for nonperturbative integrals,''
Lett Math Phys (2015) 105: 1605
[arXiv:1206.5319v4 [math-ph]].


\bibitem{Zee:2003mt}
A.~Zee,
``Quantum Field Theory in a Nutshell,''
Princeton University Press, 2010, second edition. 




\bibitem{Chiaffrino:2020akd}
C.~Chiaffrino, O.~Hohm and A.~F.~Pinto,
``Gauge Invariant Perturbation Theory via Homotopy Transfer,''
JHEP \textbf{05}, 236 (2021)
doi:10.1007/JHEP05(2021)236
[arXiv:2012.12249 [hep-th]].

\bibitem{Hall}
B.C.~Hall,
``The range of the heat kernel,''
The Ubiquitous Heat Kernel, edited by Jay Jorgensen and Lynne Walling, AMS 2006, pp. 203-231
[arXiv:math/0409308v2 [math.DG]].


\bibitem{Unruh:1976db}
W.~G.~Unruh,
``Notes on black hole evaporation,''
Phys. Rev. D \textbf{14}, 870 (1976)
doi:10.1103/PhysRevD.14.870



\bibitem{Mukhanov:2007zz}
V.~Mukhanov and S.~Winitzki,
``Introduction to Quantum Effects in Gravity,''
Cambridge University Press, 2007. 




  \end{thebibliography}
\end{document}